\documentclass[aps,prl,superscriptaddress,floatfix,preprint]{revtex4}

\usepackage{graphicx}	% Include figure files
\newcommand{\beq}{\begin{equation}}
\newcommand{\eeq}{\end{equation}}
\newcommand{\beqa}{\begin{eqnarray}}
\newcommand{\eeqa}{\end{eqnarray}}
\newcommand{\bpr}{\begin{problem}}
\newcommand{\epr}{\end{problem}}
\newcommand{\bcent}{\begin{center}}
\newcommand{\ecent}{\end{center}}
\newcommand{\bfig}{\begin{figure}}
\newcommand{\efig}{\end{figure}}
\newcommand{\bpc}{\begin{picture}}
\newcommand{\epc}{\end{picture}}
\newcommand{\barr}{\begin{array}}
\newcommand{\earr}{\end{array}}
\newcommand{\bitm}{\begin{itemize}}
\newcommand{\eitm}{\end{itemize}}
\newcommand{\bright}{\begin{flushright}}
\newcommand{\eright}{\end{flushright}}
\newcommand{\bminip}{\begin{minipage}}
\newcommand{\eminip}{\end{minipage}}
\newcommand{\btab}{\begin{tabular}}
\newcommand{\etab}{\end{tabular}}

\newcommand{\nnb}{\nonumber}
\newcommand{\reflef}{(\ref}
\newcommand{\MP}{M_{\rm P}}

\newcommand{\hiroshima}{Graduate School of Science, Hiroshima University, Kagamiyama, Higashi-Hiroshima 739-8526, Japan}

\newcommand{\izest}{International Center for Zetta-Exawatt Science and Technology, Ecole Polytechnique, Route de Saclay, Palaiseau, F-91128, France}

\begin{document}

%Title of paper
\title{
%Expected sensitivity to light Dark Matter / Dark Energy candidates 
%by searching for four-wave mixing of high-intensity lasers in the vacuum
Sensitivity to Dark Energy candidates by searching for four-wave mixing 
of high-intensity lasers in the vacuum
}

\author{Kensuke Homma} \affiliation{\hiroshima}\affiliation{\izest}

\date{\today}
%\date{August 31, 2012}

\begin{abstract}
Theoretical challenges to understand Dark Matter and Dark Energy suggest
the existence of low-mass and weakly coupling fields in the universe.
The quasi-parallel photon-photon collision system (QPS) can provide chances 
to probe the resonant production of these light dark fields 
and the induced decay by the coherent nature of laser fields simultaneously.
By focusing high-intensity lasers with different colors in the vacuum,
new colors emerge as the signature of the interaction. 
Because four photons in the initial and final states interplay 
via the dark field exchange, this process is analogous to four-wave mixing 
in quantum optics, where the frequency sum and difference among
the incident three waves generate the fourth wave with a new frequency 
via the nonlinear property of crystals.
The interaction rate of the four-wave mixing process
has the cubic dependence on the intensity of each wave. Therefore, if
high-intensity laser fields are given, the sensitivity to the weakly coupling 
of dark fields to photons rapidly increases over 
the wide mass range below sub-eV. 
Based on the experimentally measurable photon energies and the linear 
polarization states, we formulate the relation between the accessible 
mass-coupling domains and the high-intensity laser parameters, 
where the effects of the finite spectrum width of pulse lasers are taken 
into account. The expected sensitivity suggests that we have a potential 
to explore interactions at the Super-Planckian coupling strength 
in the sub-eV mass range, if the cutting-edge laser technologies are
properly combined.
\end{abstract}
\pacs{04.50.Kd, 04.80.Cc, 14.80.Va}
%04.50.Kd Modified theories of gravity
%04.80.Cc Experimental tests of gravitational theories  
%14.80.Va Axion

\keywords{}
\maketitle
%
% intro
%
\section{Introduction}
Ordinary matter occupies only $\sim 4$\% of the total energy density of 
the universe.  The remaining energies are supposed to be occupied 
by Dark Matter (DM) $\sim 23$\%
and Dark Energy (DE) $\sim 73$\%~\cite{WMAP}.
In addition to the astronomical observations,
directly probing these dark components in terrestrial laboratory experiments
has crucial roles to provide different insights into the true characters 
of the universe or the structure of the vacuum.

Light (pseudo)scalar fields are now indispensable theoretical tools to try to 
interpret the cosmological constant $\Lambda$ based on the DE scenario
\cite{DEreview}.
In reduced Planckian units with 
$c=\hbar =M_P (=(8\pi G)^{-1/2}\sim 10^{27}{\rm eV})=1$,
the observed $\Lambda \sim 10^{-120}$ is extremely small compared to
the theoretically natural scale $\Lambda \sim 1$.
There is a variety of theoretical models in the market. 
In order for a DE model to be falsifiable by laboratory experiments, 
we clarify following minimum requirements on the model:
\begin{itemize}
\item solves {\it the fine-tuning problem}; how to realize such an extremely 
small $\Lambda$,
\item solves {\it the coincidence problem}; why the energy density coincides with 
the matter density only at once at present so accidentally among the long history 
of the universe,
\item predicts the field-matter coupling strength,
the mass scale of the exchanged field, and
the measurable dynamical effect, {\it e.g.}, the force-range.
%\item no contradiction with the current observations or experimental limits.
\end{itemize}

For instance, quintessence approaches~\cite{DEreview} are
designed to resolve the {\it fine-tuning} and {\it coincidence problems}
by introducing decaying behavior of $\Lambda$ based on a potential of a scalar field.
However, the potential forms are rather phenomenologically introduced.
% missing the guiding fundamental principle. 
In the similar course, the scalar-tensor theory with 
$\Lambda$ ($STT\Lambda$)~\cite{STTL}, on the other hand,
is grounded upon the conformal transformation and the frame of observations, 
which gives several testable predictions. Therefore,
$STT\Lambda$ is one of the DE models satisfying the above requirements simultaneously.
The most significant prediction of $STT\Lambda$ is the decaying behavior of 
$\Lambda \propto t^{-2}$ as a function of time $t$. 
The present age of the universe is $t_0  \sim 1.37 \times 10^{10}$ year corresponding to
$t_0 \sim 10^{60.2}$ in reduced Planckian units.
Thus, the observed $\Lambda$ is naturally understood by the overall decaying behavior,
though expecting short-term fluctuations from the dominant behavior~\cite{STTtrap}. 
The decaying behavior depends on the conformal frame 
on which our observations are based. For example, redshift measurements relevant to DE
is implicitly based on the common atomic clock between distant points, 
{\it i.e.}, on the common elementary particle masses.  
In order to realize constancy of particle masses,
a consistent conformal frame must be favored on which the gravitational
constant $G$ looks constant, the expansion rate of the universe is consistent 
with the observation, and $\Lambda$ decays as $t^{-2}$ by keeping particle 
masses constant~\cite{FujiiShort}. 
The choice of a conformal frame unavoidably associates a massless 
Nambu-Goldstone (NG) boson, because it breaks scale invariance 
(global conformal symmetry), which is also known as dilaton. 
In contrast to the well-known Brans-Dicke model~\cite{BD}, 
a kind of scalar-tensor theories, the requirement of constancy of particle 
masses results in coupling of the scalar field with matter, 
{\it i.e.}, violation of Weak Equivalence Principle only via 
quantum anomaly coupling~\cite{STTL}.
Due to this coupling to matter fields, $STT\Lambda$ uniquely predicts 
an extremely low-mass scalar field as a pseudo NG boson via the explicit
symmetry breaking by the quantum loop effect in the self-energy.
This is similar to a massive pion as a theoretical descendant of 
an originally massless NG pseudoscalar boson associated with 
chiral symmetry.
The scalar field couples with other matter fields basically as weakly as gravity.
The mass scale $m_{\phi}$ based on the simple one-loop diagram 
in which the light quarks and leptons with a typical mass $m_{\rm q}\sim {\rm MeV}$ 
couple to the scalar field with the gravitational coupling with the strength 
$\sim \MP^{-1}$ is given by
%%%%%%%%%%%
\beq
m_\phi^2 \sim \frac{m_{\rm q}^2 M_{\rm ssb}^2}{M_{\rm P}^2}\sim (10^{-9}{\rm eV})^2,
\label{mass1}
\eeq
where the effective cutoff coming from the
super-symmetry-breaking mass-scale $ M_{\rm ssb} \sim {\rm TeV}$
is assumed, though allowing a latitude of several orders of magnitude,
if $ M_{\rm ssb}$ is higher than the conventional TeV scale~\cite{STTL}.
We note that the uncertainty on the mass range in the DE scenarios is quite large. 
Quintessence-based scenarios typically argue that the mass is determined 
from the second derivative of the assumed almost flat potential 
resulting in $m \sim 10^{-33}$~eV~\footnote{
It is rather difficult to interpret the mass scale as the actual particle mass, 
because there are no local minima in the exponentially damping flat potential form.
}.
On the other hand, $\Lambda \sim$ (meV)${}^4$ in natural units intuitively
leads models based on the particle picture\cite{QA,Chameleon,deVega} assuming 
the mass scale in the meV range via rather complicated assumptions.
For example, the axion inspired models~\cite{QA} share the similarity to $STT\Lambda$ 
by introducing the concept of pseudo NG boson driven by the two dominant scales, 
$M_P$ and a scale of the assumed symmetry breaking
at a lower energy than $M_P$
~\footnote{
These models, however, tend to fall into a kind of fine-tuning
via complicated theoretical assumptions by respecting the energy density 
at present too much.
}.

The finite mass of the scalar field in $STT\Lambda$ causes non-Newtonian force 
\cite{YFujii} via the Yukawa potential, a.k.a. fifth force.
The inverse of Eq.(\ref{mass1}) gives a finite range corresponding to $\sim 100$~m of 
the force mediated by the exchange of a quantum $\phi$ between local objects.
This is an entirely different aspect from its way of a cosmological fluid
in accelerating the universe. The force-range has not been explicit 
in the other theoretical models. This unique aspect triggered the past experimental 
efforts to measure deviations from the Newtonian potential between massive 
test bodies~\cite{FifthExp} in different contexts from DE at that time.
These measurements, however, accompany large systematic uncertainties 
due to the uncontrollable macroscopic probes. As an alternative approach, 
we have proposed to utilize the nature of 
high-intensity laser fields toward laboratory search for the scalar field 
predicted by $STT\Lambda$~\cite{DEptp} as an ultimate goal of laboratory 
experiments.

Furthermore, low-mass and weakly coupling fields are also predicted
in the contexts of particle physics with the solid foundation.
For example, axion, the pseudoscalar field is proposed as a 
NG boson associated with the global Peccei and Quinn symmetry breaking~\cite{PQ} 
to naturally maintain the $CP$ conserving nature of the QCD Lagrangian. 
Axion and invisible axion-like fields have been intensively investigated by
astrophysical objects as well as laboratory experiments~\cite{AxionReview}.
Some of them may become cold dark matter candidates,
if the mass and the coupling to matter are within the proper range.
Such fields may also leave observational isocurvature fluctuations,
if the symmetry breaking occurs during the inflation phase of 
the early universe~\cite{CDM}.
If we can anticipate that the experimental sensitivity reaches the gravitational 
coupling strength, the detection of such cold matter candidates with much 
stronger couplings to matter naturally comes into view as the preliminary 
step toward the ultimate goal.

We, therefore, generalized the principle of the measurement to 
search for both scalar and pseudoscalar fields in a model independent 
way as much as possible~\cite{DEapb}.
As amplitudes of laser fields are increased, 
we can improve the sensitivity to weakly coupling low-mass fields
predicted by any types of theories, as long as the coupling to photons is expected.
The proposed method can be regarded as a kind of particle colliders 
attempting to produce extremely light resonance states.
The mass range of interest is, however,  for instance, much lower than that 
of Higgs-like boson produced at Large Hadron Collider (LHC) by more than ten 
orders of magnitude.
In the proposed method, following two key ingredients to enhance 
the sensitivity are included. 

%The purpose of this paper is to incorporate realistic aspects of
%short laser pulses with the specification of linear polarization states
%and to formulate the relation
%between the laser parameters and achievable sensitivity to
%the mass-coupling relation. In particular, the finite spectrum widths of 
%laser fields are taken into account. 

\begin{figure}
\bcent
\includegraphics[width=12.0cm]{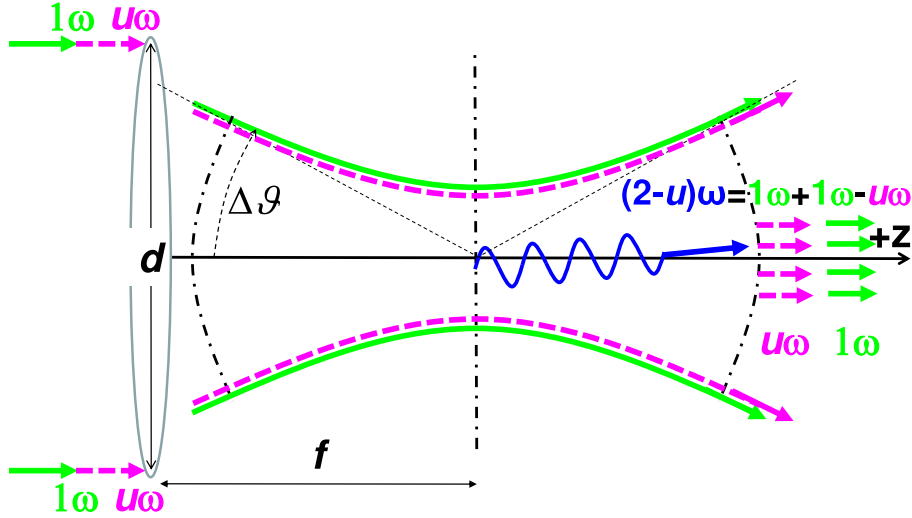}
\caption{
Quasi parallel colliding system (QPS) between two incident photons
out of a focused laser beam with the focal length $f$,
the beam diameter $d$, and the upper range of incident angles
$\Delta\vartheta$ determined by geometric optics.
The signature $(2-u)\omega$ is produced via
the four-wave mixing process,
$1\omega+1\omega\rightarrow(2-u)\omega+u\omega$ with $0<u<1$
by mixing two waves with different frequencies $1\omega$ and $u\omega$ 
in advance at the incidence.
}
\label{Fig1}
\ecent
\end{figure}

The first ingredient is the introduction of the quasi-parallel 
photon-photon collision system (QPS) as illustrated in Fig.\ref{Fig1}. 
This is considered to realize the center-of-mass system (CMS) 
energy as low as possible between colliding two laser photons 
for the production of a low-mass field as a resonance state
without lowering the incident photon energy below optical frequency.
The head-on collision in CMS corresponds to the case $\vartheta=\pi/2$
in Fig.\ref{Fig3}. The quasi-parallel system can be obtained by introducing
a Lorentz boost of the head-on collision into the perpendicular direction 
with respect to the incident direction in CMS. 
The CMS energy $E_{cms}$ is then expressed as 
$E_{cms} = 2\omega\sin\vartheta$ with the incident
energy of photon $\omega$.
If a small $\vartheta$ is realized in the laboratory frame, it provides
an extremely low CMS energy. This is the essence of the introduction
of QPS. Moreover, fortunately thanks to the strong Lorentz boost,
QPS provides frequency-shifted photons
as the decay product of the resonance state, 
which becomes a distinct observational signal as the
indication of the interaction. However, as we discuss in the
next section briefly, the resonance point cannot be
directly captured due to the extremely narrow resonance by the weakly
coupling compared with the momentum uncertainty of incident photons in QPS.
This situation requires an averaging process of the cross section
over the possible uncertainty of the CMS energy in QPS.
By this averaging, the non-negligible effect of the narrow resonance
is enhanced by the square of the inverse coupling compared with the case 
where no resonance state is contained. If the coupling to two photons
is proportional to $1/M_P$, the huge enhancement by $M^2_P$ is expected.

\begin{figure}
\bcent
\includegraphics[width=12.0cm]{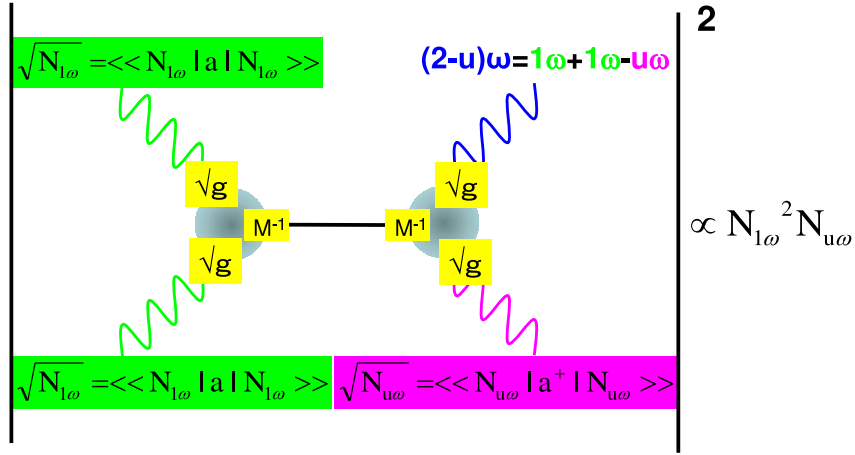}
\caption{
Four-wave mixing process in the vacuum.
The state $|N\rangle\rangle \equiv
e^{-N/2} \sum_{n=0}^{\infty} \frac{N^{n/2}}{\sqrt{n!}} |n\rangle
$ 
with $|n\rangle$ being the degenerate state of $n$ photons
refers to the coherent states~\cite{Glauber} 
with the average number of photons $N$.
The coherent states are distinguished
with photon frequencies specified by the individual subscripts.
The probability to create $(2-u)\omega$ with $0<u<1$ is expected
to have the cubic dependence on $N$, because expectation
values of creation and annihilation operators associated with the individual
photon legs become $\sqrt{N}$ in the scattering amplitude~\cite{DEptp}.
The effective dark field - two photon coupling $gM^{-1}$ for
each vertex corresponds to the coupling in the effective
interaction Lagrangian defined in Eq.(\ref{eq_phisigma}).
}
\label{Fig2}
\ecent
\end{figure}

The second ingredient is the enhancement by the coherent nature of 
laser fields~\cite{Glauber} or the degenerate nature of Bosonic particles
as illustrated in Fig.\ref{Fig2}.
This Bosonic nature of the laser beam is fundamentally important,
because we can induce the decay of the produced resonance state
into a specific momentum space as the principle of the laser amplification 
itself utilizes that nature.
We propose to use different frequencies between the production
and inducing laser beams, respectively. As shown in the figure,
the exchange of the low-mass field is interpreted as the
four-wave mixing process where three waves (the two waves are degenerate
and the one wave has a different frequency from the degenerate waves) 
are combined and the forth wave emerges with a new frequency
not included in the originally mixed laser waves.
This four-wave mixing process is well-known in quantum optics~\cite{FWM}. 
The process is already applied to generate a different color wave from
those of incident laser beams via the nonlinear atomic processes of 
crystals. In other words, the proposed method is as if
the atomic nonlinear process is replaced by the nonlinear process
of the vacuum via the low-mass field exchange.
In the context of the QED interaction, a similar approach is 
discussed~\cite{QEDchi3} and the experimental setups are proposed~\cite{QEDfwm}.
The upper limit of the photon-photon cross section is provided 
by this method~\cite{QEDlimit}.
Since each of the two photons at the first
vertex annihilates into the coherent state with $1\omega$,
while another photon at the second vertex is created from
the coherent state with $u\omega$ with $0<u<1$, 
the interaction rate to observe $(2-u)\omega$ frequency is eventually
enhanced by a factor of 
$(\sqrt{N_{1\omega}}\sqrt{N_{1\omega}}\sqrt{N_{u\omega}})^2$~\cite{DEptp},
where $N$ indicates the average number of laser photons with individual
frequency specified by the subscripts. 
This cubic dependence of the interaction rate motivates us
to make the laser energy per pulse as large as possible.

In this paper, we extend the formula discussed in the recent works
~\cite{DEptp, DEapb} and then provide the
prescription to relate the accessible mass-coupling domains
by taking an essence of realistic experimental constraint such as
a state of the multi-frequency mode with a finite frequency bandwidth
and the effect of the specification of linear polarization states,
when we attempt to apply this method to experiments 
based on pulse lasers.
The expected sensitivity to the low-mass and weakly coupling fields
is provided for anticipated high-intensity laser fields available
at laboratories over the world at present and 
in the near future~\cite{ISMD2011}. 
The result suggests that the state-of-the-art technology may 
provide access to interactions with gravitational
coupling strength and even beyond it (Super-Planckian coupling)
for relatively higher mass ranges in the sub-eV mass domain.
We emphasize that the proposed approach is a kind of Bosonic collider. 
The commonality and the distinctions from the Fermionic collider, 
for example LHC, is discussed as a concluding remark.

\section{Formulae to relate sensitivity and laser parameters}
%The purpose of this paper is to incorporate realistic aspects of
%short laser pulses and to formulate the relation
%between the laser parameters and achievable sensitivity to 
%the mass-coupling relation.
%In particular, the finite spectrum widths of laser fields are 
%taken into account. 
Let us briefly review the necessary parts for the extension
in order to consider the effects of the finite spectrum widths and 
the specification of linear polarization states of laser fields.

The effective interaction Lagrangian between two photons and
an unknown low-mass scalar or 
pseudoscalar fields $\phi$ or $\sigma$ are generalized as follows,
respectively
\beqa\label{eq_phisigma}
-L_{\phi}=gM^{-1} \frac{1}{4}F_{\mu\nu}F^{\mu\nu} \phi
\quad {\mbox or } \quad
-L_{\sigma}=gM^{-1}
\frac{1}{4}F_{\mu\nu}\tilde{F}^{\mu\nu} \sigma,
\label{mxelm_1}
\eeqa
where $M$ has the dimension of mass while $g$ being a dimensionless constant.
%If $M$ corresponds to the Planckian
%mass $M_P \sim 10^{18}$~GeV, the interaction is as weak as that of gravity
%and would have a great relevance to cosmology~\cite{STTL,DEptp}.
Depending on the allowed polarization combinations of two photons 
coupling to the dark fields, we can argue whether they are scalar-type or 
pseudoscalar-type in general as we see in Appendix in detail.
  
\begin{figure}
\bcent
\includegraphics[width=10.0cm]{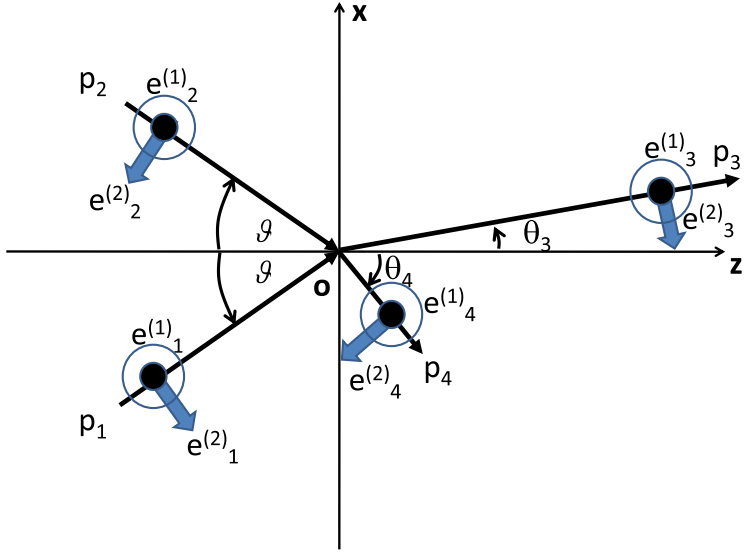}
\caption{
Definitions of kinematical variables~\cite{DEptp}.
The exact definitions of the photon momenta and the polarization vectors
can be found in Appendix.
}
\label{Fig3}
\ecent
\end{figure}

We summarize the notations and kinematics based on
the equation (2.1)-(2.3) of ~\cite{DEptp}.
We label momenta to four photons as illustrated in Fig.\ref{Fig3},
where the incident angle $\vartheta$ is assumed to be symmetric
around the $z$-axis, because we assume the
symmetric focusing as illustrated in Fig.\ref{Fig1}.
For later convenience, we introduce an arbitrary number $u$ with $0<u<1$
to re-define momenta of the final state photons as 
$\omega_4 \equiv u\omega$ and $\omega_3 \equiv (2-u)\omega$.
By this definition, we require $0 < \omega_4 < \omega_3 < 2\omega$.
We consider the case where we measure $\omega_3$ with the specified 
polarization state as the signature of the interaction.
With these definitions, energy-momentum conservation in~\cite{DEptp}
is re-expressed as
\beqa\label{eq001}
(2-u)\omega + u\omega = 2\omega
\eeqa
\beqa\label{eq002}
(2-u)\omega\cos\theta_3 + u\omega\cos\theta_4 = 2\omega\cos\vartheta
\eeqa
\beqa\label{eq003}
(2-u)\omega\sin\theta_3 = u\omega\sin\theta_4.
\eeqa
From these, we also derive the following relation
\beqa\label{eq004}
\frac{\sin\theta_3}{\sin\theta_4} = 
\frac{\sin^2\vartheta}{1-2\cos\vartheta\cos\theta_4+\cos^2\vartheta},
\eeqa
and
\beqa\label{eq005}
\omega_3 = (2-u)\omega 
= \frac{\omega\sin^2\vartheta}{1-\cos\vartheta\cos\theta_3}.
\eeqa

From (2.5) of~\cite{DEptp},
the differential cross section per solid angle of $p_3$ is expressed as
\beqa\label{eq1}
\frac{d\sigma}{d\Omega_3} =
\left( \frac{1}{8\pi\omega} \right)^2
\sin^{-4}\vartheta \left( \frac{\omega_3}{2\omega} \right) ^2
|{\cal M}_S|^2,
\eeqa
where $|{\cal M}_S|^2$ is the square of the invariant scattering
amplitude including the resonance state in the $s$-channel
with a sequence of four-photon polarization states 
$S=\beta_1\beta_2\beta_3\beta_4$ specified
by the polarization vectors $\vec{e_i}^{(\beta_i)}$ with $i=1,2,3,4$
for the photon labels, whereas $\beta=1,2$ are the kind of the
linear polarization as indicated in Fig.\ref{Fig3}.
As we discuss how to evaluate the amplitudes in Appendix, 
$S=1111,2222,1122,2211$ and $S=1212,1221,2112,2121$
give the non-vanishing invariant amplitudes
for scalar and pseudoscalar exchanges, respectively.

The enhancement by the inducing laser field labeled as $p_4$ 
is limited to the intrinsic spectrum width of the inducing laser energy 
due to energy-momentum conservation, which is defined as
\beqa\label{eq2}
\delta\omega_4 \equiv \overline{\omega_4}-\underline{\omega_4} 
               = (\delta u) \omega,
\eeqa
where $\delta u \equiv \overline{u}-\underline{u}$
with $0 < \overline{u} < 1$, $0 < \underline{u} < 1$ and
$\overline{u} > \underline{u}$.
The overline and underlines attached to the simple variables 
indicate the allowed maximum and minimum values, respectively.
This notation is used repeatedly, unless confusion occurs.
The spectrum width of $\omega_3$ is simultaneously
constrained to 
\beqa\label{eq3}
\delta\omega_3 \equiv
\overline{\omega_3}-\underline{\omega_3}
=
(2-\underline{u})\omega - (2-\overline{u})\omega
= (\delta u) \omega
\eeqa
due to energy-momentum conservation as well.
This indicates that the energy range to detect $\omega_3$ must be consistent
with $\delta\omega_4$ in order to get the full enhancement factor
by the inducing laser.

By assuming that $\delta\omega_3 = \delta\omega_4$ is realized
in an experimental setup,
let us now calculate the cross section integrated over $\delta\omega_3$
as a function of the spectrum width parameter $(\overline{u},\underline{u})$.
Eventually $\delta\omega_3$ gives the range of the integral via
the range of scattering angle $\theta_3$.
The integrated cross section over the range of
$\theta_3$ from $\underline{\theta_3}$ to $\overline{\theta_3}$ is expressed as
\beqa\label{eq4}
\overline{\sigma}
% (\underline{u},\overline{u})
=
\frac{{\cal F}_S \overline{|{\cal M}_S|^2}}{(8\pi\omega)^2 \sin^4\vartheta}
\int_{\underline{\theta_3}}^{\overline{\theta_3}} 
\left( \frac{\omega_3}{2\omega} \right)^2 \sin\theta_3 d\theta_3
\equiv
\frac{{\cal F}_S \overline{|{\cal M}_S|^2}}{(8\pi\omega)^2 }{\cal I},
\eeqa
by defining
\beqa\label{eq5}
{\cal I} \equiv
\frac{1}{\sin^4\vartheta} \int_{\underline{\theta_3}}^{\overline{\theta_3}}
\left( \frac{\omega_3}{2\omega} \right)^2 \sin\theta_3 d\theta_3
= \left[ \frac{1}{4\cos\vartheta(1-\cos\theta_3\cos\vartheta)} \right]^{\underline{\theta_3}}_{\overline{\theta_3}},
\eeqa
where $\overline{|{\cal M}_S|^2}$ denotes the averaged $|{\cal M}_S|^2$
over possible uncertainty on the incident angle $\vartheta$ as we briefly
discuss below, and ${\cal F_S}$ corresponds to the integral over 
the azimuthal degree of freedom depending on the specification of 
photon polarizations in the initial and final states. 
If the scattering amplitude has 
the axial symmetry around the $z$-axis in Fig.\ref{Fig3}, 
${\cal F_S}$ simply corresponds to $2\pi$. As we summarize
in Appendix, however, the axial asymmetries actually appear
depending on $S$ specified by experimental conditions, which
result in deviations from $2\pi$.
We then convert the variable $\cos\theta_3$ to $u$ 
based on Eq.(\ref{eq005}) from which we express 
\beqa\label{eq6}
\cos\theta_3 = \frac{1}{\cos\vartheta} 
\left( 1-\frac{\omega}{\omega_3}\sin^2\vartheta \right)
\sim 1+\frac{1}{2} \vartheta^2 \left( 1-\frac{2\omega}{\omega_3} \right)
= 1+\frac{1}{2} \vartheta^2 \left( 1-\frac{2}{2-u} \right),
\eeqa
where terms of the order higher than $\vartheta^2$ are dropped
when applied to the low-mass resonance.
From this, Eq.(\ref{eq5}) is also approximated as
\beqa\label{eq7}
{\cal I} \sim \frac{\delta u}{4\vartheta^2}.
\eeqa
%With (\ref{eq4}) and (\ref{eq7}) we can express $\overline{\sigma}$
%as a function of the spectrum width parameter $(\overline{u},\underline{u})$
%\beqa\label{eq8}
%\overline{\sigma} \sim 
%\frac{\pi}{2} \delta u (8\pi\omega)^{-2}\vartheta^{-2}
%\overline{|{\cal M}|^2}.
%\eeqa
%

We now consider the scattering amplitudes only for the case when 
low-mass fields are exchanged via the resonance states in the s-channel.
The resonance decay rate of the low-mass field with the
mass $m$ into two photons is expressed as~\cite{DEptp, DEapb}
%%%%%%%%%%
\beq
\Gamma=(16\pi)^{-1} \left( g M^{-1}\right)^2 m^3.
\label{mxelm_4a}
\eeq
As we calculate in detail in Appendix, for example,
in the case of scalar field exchange, 
the invariant amplitude in the coplanar condition where
the plane determined by $p_1$ and $p_2$ coincides with that determined by
$p_3$ and $p_4$ is expressed as
%%%%%%%%%%%
\beq
{\cal M}_S =-(g M^{-1})^2\frac{\omega^4 \left(
\cos2\vartheta -1\right)^2}{2\omega^2 \left(
\cos2\vartheta -1\right)+m^2},
\label{mxelm_7}
\eeq
where the denominator, denoted by ${\cal D}$ in the following,
is the low-mass field propagator.
We then introduce the imaginary part due to the resonance state
by the following replacement
%%%%%%%%%
\beq
m^2 \rightarrow \left( m -i\Gamma \right)^2 \approx
m^2 -2im \Gamma.
\label{mxelm_9}
\eeq
Substituting this into the denominator in Eq.~\reflef{mxelm_7}) and
expanding around $m$, we obtain
%%%%%%%%%%%
\beq
\hspace{-.1em}{\cal D}\approx -2\left( 1-\cos2\vartheta  \right) \left( \chi+ia
\right),\quad\hspace{-.7em}\mbox{with}\quad\hspace{-.7em} \chi =\omega^2 -\omega_r^2,
\label{mxelm_10}
\eeq
where
%%%%%%%%%%%%
\beq
\omega_r^2 =\frac{m^2/2}{1-\cos 2\vartheta },\quad
a=\frac{m \Gamma}{1-\cos 2\vartheta}.
\label{mxelm_12}
\eeq
From Eq.~\reflef{mxelm_4a}) and \reflef{mxelm_12}),  $a$ is also 
expressed as
\beq\label{eq_a}
a = \frac{\omega^2_r}{8\pi}\left(\frac{g m}{M}\right)^2,
\eeq
which explicitly shows the proportionality to $M^{-2}$.
We then express the squared amplitude as
\beq
|{\cal M}_S|^2 \approx  (4\pi)^2 \frac{a^2}{\chi^2+a^2}.
\label{mxelm_13}
\eeq
Theoretically if we take the limit of $\omega \rightarrow \omega_r$,
$|{\cal M}_S|^2 \rightarrow (4\pi)^2$ is realized from (\ref{mxelm_13}).
This is independent of the smallness of $a \propto M^{-2}$.
Meanwhile, the off-resonance case $\chi\gg a$, $|{\cal M}_S|^2$
equivalent to Eq.(\ref{mxelm_7})
is largely suppressed due to the factor $a^2 \propto M^{-4}$ for
the case of a weakly coupling $M^{-1}$.

This is the most important feature arising from the resonance that 
overcomes the weakly coupling stemming from the large relevant mass scale 
such as $M=M_P$.
However, we are then confronted with an extremely narrow width $a$
for {\it e.g.} $gm \ll 1$~eV , $M \sim M_P=10^{27}$~eV and
$\omega_r \sim \mbox{1 eV}$.
The rescue to overcome this difficulty is the averaging process
over unavoidable uncertainties of incident angles in QPS.
Even if a single photon with a fixed frequency $\omega$ is focused by a
lens element in QPS, the wave vector around the diffraction limit 
fluctuates by the wavy nature, in other words, the beam waist
$\Delta x$ at the diffraction limit and the momentum accuracy
$\Delta p$ must
satisfy the uncertainty principle $\Delta x \Delta p \ge \hbar/2$.
Therefore, we need to distinguish between the theoretically specified
momenta and the physically specified ones in QPS.

Based on Eq.(40),(41), and (42) of~\cite{DEapb}, we express the average of
the square of the invariant amplitude over a possible uncertainty 
on the incident angle $\vartheta$, {\it i.e.}, uncertainty on
the directions of the wave vectors as
\beqa\label{eq9}
\overline{|{\cal M}_S|^2}=\int_{0}^{\pi/2}\rho(\vartheta)|{\cal M}_S|^2 d\vartheta
\sim \frac{(4\pi)^2}{2\omega^2}
\left(\frac{\vartheta_r}{\Delta\vartheta}\right)a\pi,
\eeqa
where
the resonance angle $\vartheta_r$ satisfies the resonance condition
$\omega^2 = \omega_r = (m^2/2)/(1-\cos2\vartheta_r)$,
and we plugged the following simplest angular distribution 
function $\rho$~\footnote
{This step-like distribution might be over 
simplified, however, it is useful to provide a conservative 
sensitivity. This function must be determined by the individual 
experimental setup based on the actual profile measurement 
in QPS eventually.} 
into Eq.(\ref{eq9}):
\beqa\label{eq10}
\rho(\vartheta) = \left\{
\begin{array}{ll}
1/\Delta\vartheta & \quad \mbox{for $0 < \vartheta \le \Delta\vartheta$} \\
0 & \quad \mbox{for $\Delta\vartheta < \vartheta \le \pi/2$}
\end{array}
\right\},
\eeqa
which is normalized to the physically possible range $0< \vartheta \le \pi/2$.
The incident angle uncertainty
$\Delta\vartheta$ can be as large as that determined
from geometrical optics~\footnote{
The Gaussian beam waist $w$ at the diffraction limit is known as 
$w=\frac{f\lambda}{\pi(d/2)}$~\cite{Yariv}. The momentum uncertainty
of an incident photon at the waist is therefore 
$\Delta p \sim \frac{\hbar/2}{2w}$ from the uncertainty principle.
With the laser photon momentum $p=\frac{h}{\lambda}$, the angle 
uncertainty at the waist is given by 
$\Delta \vartheta \sim \frac{\Delta p}{p} \sim \frac{1}{8}\frac{d/2}{f}$.
This range is somewhat smaller than that of geometric optics, however,
we rather take the classical limit as a conservative range.
This is because the interaction is not necessarily limited only around 
the waist, but may happen at any points during the propagation into the
diffraction limit, as we discuss later.
}
\beqa\label{eq15}
\Delta\vartheta \sim \frac{d}{2f},
\eeqa
with the common beam diameter $d$ and focal length $f$ 
for both the creation and inducing beams
as illustrated in Fig.\ref{Fig1}.
%For later convenience, we put the alternative expression for $a$
%\beqa\label{eq11}
%a \equiv \frac{\omega^2}{16\pi}\left(\frac{gm}{M}\right)^2.
%\eeqa
%

By substituting Eq.(\ref{eq7}), (\ref{eq9}) and (\ref{eq_a}) into
Eq.(\ref{eq4}), we express the partially integrated
averaged cross section as
%as a funcion of the spectrum width
\beqa\label{eq21}
\overline{\sigma} =
%(\underline{u},\overline{u}) =
\frac{{\cal F_S} \overline{|{\cal M}_S|^2}}{(8\pi\omega)^2 }{\cal I}
\sim 
%\frac{2\pi}{(8\pi\omega)^2 } \cdot
%\frac{(4\pi)^2}{2\omega^2}
%\left(\frac{\vartheta_r}{\Delta\vartheta}\right)a\pi \cdot
%\left(\frac{\overline{u}-\underline{u}}{4\vartheta^2}\right)
%=
\frac{{\cal F_S}}{(8\pi\omega)^2 } \cdot
\frac{(4\pi)^2}{2\omega^2}
\left(\frac{\vartheta_r}{\Delta\vartheta}\right)
\frac{\omega^2}{16\pi}\left(\frac{gm}{M}\right)^2 \pi \cdot
\frac{\delta u}{4\vartheta^2}\nnb\\
\sim
\frac{{\cal F_S}}{512\omega^2}
\frac{\delta u}{\vartheta_r  \Delta\vartheta}
\left(\frac{gm}{M}\right)^2 
=
\frac{1}{512}\left(\frac{\lambda}{2\pi}\right)^2
\frac{\delta u}{\vartheta_r  \Delta\vartheta}
{\cal F_S} \left(\frac{gm}{M}\right)^2.
\eeqa
In the averaging process, among a possible range of $\vartheta$, 
only $\vartheta \sim \vartheta_r$ effectively contributes to 
the cross section because of the narrow width $a$ in Eq.(\ref{eq_a})
for a large $M$. 
The second line in Eq.(\ref{eq21}) takes this aspect into account. 
For the last equation,
$\omega\mbox{[eV]} = 2\pi\hbar c/\lambda$ with $\hbar=c=1$ is substituted,
where $\lambda$ is the wavelength of the creation laser field.

We now express the yield of frequency-shifted photons ${\cal Y}$ 
as a function of the spectrum width parameter $(\overline{u},\underline{u})$
\beqa\label{eq12}
{\cal Y } = {\cal L} \overline{\sigma}
\eeqa
with the effective integrated luminosity ${\cal L}$ over 
propagation time of a single shot laser fields with
pulse duration time $\tau$ which is assumed to be common for both
the creation and inducing beams.
We discussed how the integrated effective luminosity 
should be defined in~\cite{DEapb}. Here we briefly review the relevant part.
The solution of the electromagnetic field
propagation in the vacuum with a Gaussian profile in the transverse plane
is well-known~\cite{Yariv}.
The transverse spatial profile of a laser field is typically
Gaussian to the first order approximation in high-intensity laser systems.
In this case, the electric field propagating along the $z$-direction
in spatial coordinates $(x,y,z)$ is expressed as
%
%%%%%%%%%%
\beqa\label{eq_Gauss}
E(x,y,z) \propto
\qquad \qquad \qquad \qquad \qquad \qquad \qquad \qquad \qquad \nnb\\
\frac{w_0}{w(z)}\exp
\left\{
-i[kz-H(z)] - r^2 \left( \frac{1}{{w(z)}^2}+\frac{ik}{2R(z)} \right)
\right\},
\eeqa
%%%%%%%%%%
%
where $k=2\pi/\lambda$, $r=\sqrt{x^2+y^2}$, $w_0$ is the minimum waist,
which cannot be smaller than $\lambda$ due to the diffraction limit, and
other definitions are as follows:
%
%%%%%%%%%%
\beqa\label{eq_wz}
{w(z)}^2 = {w_0}^2
\left(
1+\frac{z^2}{{z_R}^2}
\right),
\eeqa
\beqa\label{eq_Rz}
R = z
\left(
1+\frac{{z_R}^2}{z^2}
\right),
\eeqa
\beqa\label{eq_etaz}
H(z) = \tan^{-1}
\left(
\frac{z}{z_R}
\right),
\eeqa
\beqa\label{eq_zr}
z_R \equiv \frac{\pi{w_0}^2}{\lambda}.
\eeqa
The transverse beam size of the focused Gaussian laser beam
is minimized at the beam waist and then expands beyond
the focal point where interactions among out-going photons are prohibited
by the condition that two photons propagate into opposite directions in CMS. 
On the other hand, the exchange of a low-mass field may take place anywhere 
within the volume defined by the transverse area of the Gaussian laser times 
the focal length $f$ before reaching the focal point (see Fig.\ref{Fig1}).

Given the Gaussian laser parameters above, the effective integrated
luminosity ${\cal L}$ over the propagation volume of a laser pulse 
can be defined as follows~\cite{DEapb}.
%In this paper, we consider only the case for a short pulse laser, 
%where the force-range is assumed to be larger than the spatial width of 
%a laser pulse along the $z$-axis, namely, $\hbar /(mc) > c\tau$ for 
%$mc^2<1$~eV restoring the physical dimensions of $\hbar$ and $c$.
%
At an instant, the interaction is limited within a region over $c\tau$
where the average number of photons $N_c$ and $N_i$ are available for
creation and inducing processes, respectively.
Luminosity at a point $z$ integrated over pulse duration $\tau$ is 
expressed as
\beqa\label{eq_Linstant}
{\cal L}(z) = \frac{I(N_c, N_i)}{\pi w^2(z)}
= \frac{I(N_c, N_i)}{\pi w^2_0}\frac{z^2_R}{z^2+z^2_R}
\eeqa
where
$I(N_c, N_i)$ denotes a dimensionless intensity depending dominantly
on the average number of creation and inducing photons, $N_c$ and $N_i$,
respectively within duration time $\tau$.
The expression $w^2(z)$ in Eq.~(\ref{eq_wz}) is substituted.
During the propagation over the focal length $f$, the effective
number of the interacting regions or the number of virtual bunches $b$ is
expressed as $b=f/(c\tau)$.Therefore,
the effective integrated luminosity ${\cal L}$ over pulse propagation
time averaged over the focal length $f$ is finally expressed as
\beqa\label{eq_Lbar}
{\cal L} = b \int^f_0 f^{-1} {\cal L}(z) dz =
\frac{I(N_c, N_i)}{c\tau \lambda} \tan^{-1}\left(\frac{f}{z_R}\right).
\eeqa

In the case of charged particles or Fermionic beams, 
the dimensionless intensity $I$ corresponds to the square of the number of 
particles per bunch, which is the combinatorics 
to take two Fermions from individual 
colliding beam bunches. On the other hand, in the case of four-wave mixing, 
all photons are in the quantum coherent states with
the inducing nature resulting in the cubic dependence as we discussed. 
By taking this aspect into account, we define the dimensionless
intensity included in ${\cal L}$ as follows
\beqa\label{eq13}
I(N_c, N_i) \equiv C_{mb} N^2_c {\cal A} N_i,
\eeqa
where
${C_{mb}}$ is a factor to consider combinatorics
for the choice of two photons in the creation beam and
one photon in the inducing beam by extending the argument
for the single-frequency mode~\cite{DEptp} to the multi-frequency mode
as discussed below,
and ${\cal A}$ is an acceptance factor for the inducing photons
to satisfy energy-momentum conservation in the final state.

The acceptance factor ${\cal A}$ is introduced because
the process occurs only in a small portion $\delta\theta_4$ of the
entire angular spectrum with the whole strength $N_i$
distributed over the total range of $\Delta\theta_4$, hence;
%
%We now discuss the acceptance factor ${\cal A}$ of the inducing beam.
%This factor originates from a situation where all of the inducing beam
%$N_i$ is not useful for the enhancement, 
%because the range, $\delta\theta_4$, determined by the intrinsic 
%spectrum width of the inducing laser beam is narrower than the prescribed 
%angle range determined by the focusing parameter based on
%geometrical optics.
%
%Therefore, this factor is defined as
\beqa\label{eq15}
{\cal A} \equiv \frac{\delta\theta_4}{\Delta\theta_4}
= \frac{\delta\theta_4}{\Delta\vartheta},
\eeqa
which is much smaller than unity.
Here $\Delta\theta_4$ is further assumed to be common with that of the
creation beam $\Delta\vartheta$ by sharing the same optics
as that of the creation beam.
The $\delta\theta_4$ is constrained by the spectrum width
of $\omega_4 = u\omega$, therefore, is described as
a function of the spectrum width parameter $(\overline{u},\underline{u})$
as follows
\beqa\label{eq16}
\delta\theta_4 =
\left(
\sqrt{\frac{2-\underline{u}}{\underline{u}}}-
\sqrt{\frac{2-\overline{u}}{\overline{u}}}
\right)\vartheta \equiv {\cal U} \vartheta.
\eeqa
This relation is obtained from energy-momentum conservation as follows.
Equation (\ref{eq003}) gives the ratio ${\cal R}$ defined by
\beqa\label{eq17}
{\cal R} \equiv
\frac{\sin\theta_3}{\sin\theta_4} = 
\frac{\omega_4}{\omega_3} = \frac{u}{2-u}.
\eeqa
By equating Eq.(\ref{eq17}) and (\ref{eq004}), we obtain
\beqa\label{eq19}
\cos\theta_4 = \frac{{\cal R}(1+\cos^2\vartheta)-
\sin^2\vartheta}{2{\cal R}\cos\vartheta}.
\eeqa
Neglecting higher order terms more than $\vartheta^2$, 
we approximate $\theta_4$ as
\beqa\label{eq20}
\theta_4 \sim \sqrt{\frac{1}{{\cal R}}}\vartheta
= \sqrt{\frac{2-u}{u}}\vartheta.
\eeqa
For a low-mass case $m \sim 2\vartheta\omega$
with $\vartheta \ll 1$, $\theta_4$ is also small 
via $\theta_4 \sim {\cal R}^{-1/2}\vartheta$ from Eq.(\ref{eq20}).
Eventually the emission angle of the signal, $\theta_3$ also becomes small
via $\theta_3 \sim {\cal R}^{1/2}\vartheta$ from 
Eq.(\ref{eq17}) and (\ref{eq20}).
This is the reason why the creation and inducing beams
are all aligned into the same optical axis $z$ as illustrated
in Fig.\ref{Fig1}, by which a chance to enhance the inducing process is
maximized. We note here that we have only to search for $(2-u)\omega$
as the signature of the four-wave mixing process
without measuring the emission angle $\theta_3$ directly.

We then consider the effect of multi-frequency mode of a short pulse laser
via the following argument 
on the combinatorics factor $C_{mb}$ in the mixing process.
This is physically unavoidable, 
because a pulse laser with a short time duration must
contain the corresponding energy uncertainty in principle.
We assume that photons of a creation laser pulse in a state of 
multi-frequency mode can be uniformly divided into $n_c$ frequency bins, 
namely, forming an uniform frequency density within the frequency bandwidth 
of the creation laser pulse, 
each of which forms a coherent state of the individual frequency.
We also assume $n_i$ for the inducing laser pulse as well. 
All frequencies as a result of the possible mixing between creation and 
inducing frequencies must be detectable by an experimental setup 
in order for the following argument to be valid. 
We also assume all the frequency modes share the common focusing length
and beam diameter independent of the wavelengths as illustrated 
in Fig.\ref{Fig1}.
In this model case, the dimensionless intensity in Eq.(\ref{eq13}) 
included in the luminosity definition 
is expected to be \beqa\label{eq14}
I \propto
\frac{1}{2} n^2_c
\left(\sqrt{\frac{N_c}{n_c}}
      \sqrt{\frac{N_c}{n_c}}\right)^2 
n_i
\left(\sqrt{\frac{N_i}{n_i}}\right)^2 
=
\frac{1}{2}{N_c}^2{N_i},
\eeqa
where 
square roots are enhancement factors due to the coherent states of
individual frequency bins,
$\frac{1}{2} n^2_c$ corresponds to combinatorics to choose
initial two photons within $n_c$ frequency bins of a creation laser pulse 
by allowing to choose two photons even within the same frequency bin, while
$n_i$ is the degree of freedom to chose one frequency bin out of $n_i$ bins.
We thus find $C_{mb} = \frac{1}{2}$ which is the same dimensionless
intensity as discussed in~\cite{DEapb} after all.

The effect of the multi-frequency mode is
not seen directly in terms of the combinatorics, however, 
we put a note that choosing two photons $\omega_1$ and 
$\omega_2$ out of $n_c$ bins implies that the incident 
two photon energies could be different, which deviates
from the simplest assumption of the symmetric angle of incidence with equal
photon energies as illustrated in Fig.\ref{Fig1}.
However, we can always find a reference frame by a Lorentz boost
which exactly satisfies the symmetric conditions by considering 
the inverse process of the massive particle decay into 
two photons starting from the rest frame of that particle,
because the interaction is enhanced only when the resonance condition
$m \sim 2\vartheta\omega$ is fulfilled.
Instead, $p_3$ in QPS must contain fluctuations by this boost effect.
Therefore, as long as the experimental coverage on $p_3$ with 
respect to the additional fluctuations is broad enough, 
the integrated interaction rate over 
the detector acceptance is unchanged from the case with single mode lasers.
In order to estimate the necessary coverage of $\omega_3$ approximately,
we may re-define $\omega \equiv (\omega_1+\omega_2)/2$ as the
averaged value between arbitrarily selected two incident photon energies
by taking the approximation where the transverse momentum of the 
produced massive particle is negligibly small 
compared to the longitudinal momentum of the produced particle in QPS,
and all equations in Eq.(\ref{eq001})-(\ref{eq004}) are restored,
because they are simply scaled by the newly defined $\omega$. 
Therefore, the arguments so far are approximately valid even in the
case of the multi-mode lasers.
Suppose that the bandwidth of the creation beam and inducing beams
are defined, respectively as
\beqa\label{eq14a}
\langle\omega\rangle-\Delta\omega \le \omega 
\le \langle\omega\rangle +\Delta\omega
\eeqa
\beqa\label{eq14b}
\langle\omega_4\rangle-\Delta\omega_4 \le \omega_4 
\le \langle\omega_4\rangle +\Delta\omega_4,
\eeqa
where $\langle\quad\rangle$ denotes an average of each frequency 
distribution, while
$\Delta$ indicates half of the full bandwidth of each frequency.
According to the last part of Eq.(\ref{eq2}), $\delta\omega_4$ varies 
depending on an arbitrary chosen $\omega$. However, $u$ is actually defined 
with respect to $\omega$. Therefore, any chosen $\omega$ are absorbed 
into the definition of $u \equiv \omega_4/\omega$. Thus only the
first part of Eq.(\ref{eq2}) becomes essentially relevant, which is
determined by the intrinsic character of the prescribed inducing laser beam 
independent of the creation laser frequency $1\omega$.
Accordingly an experiment should be designed so that all the modified range of 
$\delta\omega_3$ via relation Eq.(\ref{eq2}) and (\ref{eq3}) is acceptable.
Let us introduce new notations to describe the modified range of
$\omega_3$ due to the shift of the averaged $\omega$ in order to distinguish it
from $\delta\omega_3$ intrinsically caused by
$\delta\omega_4$ via Eq.(\ref{eq2}) and (\ref{eq3}).
The modified upper and lower edges of $\omega_3$ can be defined, respectively as
\beq\label{eq14c}
\overline{\omega_3^{'}} \equiv 2\overline{\omega} - \underline{\omega_4}
= 2(\langle\omega\rangle+\Delta\omega)-(\langle\omega_4\rangle-\Delta\omega_4)
%= (2\langle\omega\rangle-\langle\omega_4\rangle)
%+ (2\Delta\omega + \Delta\omega_4)
\equiv \langle\omega_3\rangle + \Delta\omega_3
\eeq
\beq\label{eq14d}
\underline{\omega_3^{'}} \equiv 2\underline{\omega} - \overline{\omega_4}
= 2(\langle\omega\rangle-\Delta\omega)-(\langle\omega_4\rangle+\Delta\omega_4)
%= (2\langle\omega\rangle-\langle\omega_4\rangle)
%- (2\Delta\omega + \Delta\omega_4)
\equiv \langle\omega_3\rangle - \Delta\omega_3,
\eeq
with
$\langle\omega_3\rangle \equiv 2\langle\omega\rangle-\langle\omega_4\rangle$
and 
$\Delta\omega_3 \equiv 2\Delta\omega + \Delta\omega_4$.
As long as an experiment can accept $\omega_3$ in the range
$\underline{\omega_3^{'}} \leq \omega_3 \leq \overline{\omega_3^{'}}$,
the expected yield in the proceeding paragraphs is valid.
Naturally, the range of $\delta\omega_3$ via the intrinsic $\delta\omega_4$ is 
fully contained in this modified range.
In the case of the multi-frequency mode, strictly speaking, 
$u$ should be defined as 
$u \equiv \langle\omega_4\rangle/ \langle\omega\rangle$
resulting in 
$\overline{u} \equiv (\langle\omega_4\rangle+\Delta\omega_4) 
/ \langle\omega\rangle$
and 
$\underline{u} \equiv (\langle\omega_4\rangle-\Delta\omega_4) 
/ \langle\omega\rangle$.
For the entire arguments in this paper, 
$u$ and $\delta u$ are implicitly assumed to be defined 
based on the averaged frequencies.
%As a short summary of this paragraph, the effect of the
%multi-frequency mode basically does not change what we have argued
%in \cite{DEptp} dedicated for the single-frequency mode
%except the additional factor $C_{mb} = 1/2$.

We now express the yield ${\cal Y}$ by substituting
Eq.(\ref{eq13}), (\ref{eq15}) and (\ref{eq21}) into Eq.(\ref{eq12})
\beqa\label{eq22}
{\cal Y } 
%=
%\tan^{-1}\left(\frac{f}{z_R}\right)
%\frac{{C_{mb}}}{c\tau\lambda}
%{N_c}^2 \cdot \frac{{\cal U}\vartheta}{\Delta\vartheta} \cdot
%{N_i} \cdot \frac{\pi}{256}\left(\frac{\lambda}{2\pi}\right)^2
%\frac{\delta u}{\vartheta_r  \Delta\vartheta}
%\left(\frac{gm}{M}\right)^2 \nnb\\
%\sim
%\frac{\lambda}{c\tau} \cdot
%\frac{tan^{-1}\left(\frac{f}{z_R}\right)}{1024\pi (\Delta\vartheta)^2} \cdot
%{\cal U} \delta u \cdot
%\left(\frac{gm}{M}\right)^2 \cdot
%{C_{mb}}{N_c}^2 {N_i} \nnb\\
\equiv
K_0(\lambda,\tau) K_1(f, d)  K_2(\overline{u},\underline{u})
{\cal F_S} \left(\frac{gm}{M}\right)^2 {C_{mb}}{N_c}^2{N_i},
\eeqa
where 
\beq\label{eqK0}
K_0(\lambda,\tau) \equiv \frac{\lambda}{c\tau}
\eeq
indicates that shorter pulse duration time 
$c\tau \rightarrow \lambda$ has the maximum gain on the yield,
\beq\label{eqK1}
K_1(f,d) \equiv
\frac{1}{2048\pi^2 (\Delta\vartheta)^2} \tan^{-1}\left(\frac{f}{z_R}\right)
\sim \frac{1}{512\pi^2} \left(\frac{f}{d}\right)^2 \tan^{-1}
\left(\frac{\pi d^2}{4f\lambda}\right)
\eeq
with $w_0 = f\lambda/(\pi(d/2))$ for $z_R = \pi w^2_0/\lambda$~\cite{Yariv}
is the parameter relevant to only optics, and
\beq\label{eqK2}
K_2(\overline{u},\underline{u}) \equiv {\cal U} \delta u =
\left(
\sqrt{\frac{2-\underline{u}}{\underline{u}}}-
\sqrt{\frac{2-\overline{u}}{\overline{u}}}
\right) (\overline{u}-\underline{u})
\eeq
is the laser spectrum width parameter determined
from the spectrum width of the inducing laser field, $\delta\omega_4$.

From Eq.(\ref{eq22}) we finally obtain the expression for the coupling
parameter $g/M$ to discuss the sensitivity as a function of $m$
for a given experimental parameters via the following equation,
\beqa\label{eq23}
\frac{g}{M} = 
m^{-1}
\sqrt{\frac{{\cal Y}}
{K_0 K_1 K_2{\cal F_S} {C_{mb}}{N_c}^2{N_i}}}.
\eeqa
For convenience to design experiments,
we also consider the case where duration times are
not equal between the creation beam $\tau_c$ and the
inducing beam $\tau_i$. Since the yield is increased by the
quadratic dependence on the creation beam intensity, it is natural 
to realize the case $\tau = \tau_c \le \tau_i$ in experiments.
If this is the case, the accessible coupling is re-expressed as
\beqa\label{eq24}
\frac{g}{M} = 
m^{-1}
\sqrt{\frac{{\cal Y}}
{K_0 K_1 K_2{\cal F_S} {C_{mb}}{N_c}^2{N_i} (\tau_c/\tau_i)}}.
\eeqa

\section{Expected sensitivity}
First we summarize the key control parameters or experimentally adjustable 
knobs based on the arguments in the previous sections.
The resonance condition is satisfied if $m \sim 2\vartheta\omega$.
However, instead of hitting the resonance point directly,
our approach is to take the average of the squared scattering amplitude
over the possible incident angle uncertainty $\Delta\vartheta$ in QPS by the
focused laser beams.
Changing the focal length introduces different $\Delta\vartheta$, {\it i.e.},
the different range of the angular integral for the averaging.
If a resonance peak is contained in that range,
the resonance effect appears as the integrated result.
The basic strategy of this proposal is therefore
to change the focal length, attempting to search for the appearance
of four-wave mixing photons, which approximately gives a mass range
via the relation $m \le 2\omega\Delta\vartheta$. Thus,
this knob provides variations along the mass axis, $m$.
On the other hand, the four-wave mixing yield is enhanced by the cubic
product of the laser intensities. Therefore, changing laser intensities 
gives large variations along the coupling axis, $g/M$. 
If a significant signature is found, we can
localize the domain in the $(m, g/M)$ plane by adjusting these two knobs.

In addition, there are knobs on the polarization states of laser fields. 
As we discuss in detail in Appendix,
depending on the types of exchanged fields, different polarization 
correlations are expected between the two photons in the
initial and final states.
In the case of the scalar field exchange which is the
first of Eq.(\ref{eq_phisigma}),
the possible linear polarization states in the four-wave mixing process
are expressed as follows:
\begin{eqnarray}\label{Aeq4}
\omega\{1\}+\omega\{1\} \rightarrow (2-u)\omega\{1\} + u\omega\{1\} \\ \nnb
\omega\{1\}+\omega\{1\} \rightarrow (2-u)\omega\{2\} + u\omega\{2\},
\end{eqnarray}
where 
photon energies from the initial to final state are denoted with
the linear polarization states
\{1\} and \{2\} which are orthogonal each other. 
On the other hand, in the case of the pseudoscalar field exchange
with the second of Eq.(\ref{eq_phisigma}),
the possible linear polarization states are expressed as:
\begin{eqnarray}\label{Aeq5}
\omega\{1\}+\omega\{2\} \rightarrow (2-u)\omega\{2\} + u\omega\{1\} \\ \nnb
\omega\{1\}+\omega\{2\} \rightarrow (2-u)\omega\{1\} + u\omega\{2\}.
\end{eqnarray}
By choosing physically allowed combinations of the linear polarizations, 
we can distinguish the types of exchanged fields,
while we can estimate the background processes by requiring the
false combinations on purpose.

%Facilities:
%ELI\cite{ELI}.
%
%IZEST (International Center for Zetta- and Exawatt Science and Technology)
%dedicates its laser (PETAL with $> 3$kJ per beam )
%to the laser electron acceleration project\cite{IZEST-HP}.
%
%ICAN (International Coherent Amplification Network)\cite{ICAN-HP}
%has been established to nurture this technology toward the application
%of the laser-driven collider (and other large averaged power applications).
%
%a MJ laser such as the LMJ laser~\cite{LeGarrec}. 
%
%NIF
%
Let us now briefly review some of major high-intensity laser facilities 
in the world including on-going projects which can provide more than 100~J per pulse. 
There are typically two classes of laser systems to achieve high-intensity:
a moderate pulse energy per tens of fs short duration and a large 
pulse energy per several ns duration.
The established choices of the laser technology for the former and 
latter classes are Titan:sapphire-based lasers and Nd:glass-based lasers 
typically dedicated for laser fusion studies, respectively.
We note that energy per shot is more important than 
pulse duration for the proposed method, because the interaction rate is 
cubic to the numbers of photons, while it is inversely proportional to
pulse duration. On the other hand, the typical repetition rates for such 
high-energy pulse lasers are currently limited to at most every minute and every 
several hours for the former and latter classes, respectively.
The frontiers of the former class are
{\it VULCAN 10PW} 300J/30fs\cite{VULCAN10PW}, {\it APOLLON} 150J/15fs\cite{APOLLON},
and what is prepared for the Extreme Light Infrastructure (ELI) project by combining 
the 20 APOLLON-type lasers\cite{ELI}.
The examples of the latter class are
{\it GEKIKO-XII} 0.1-1kJ combining 12 beam lines~\cite{OSAKA},
{\it VULCAN} 2.6kJ combining 8 beam lines~\cite{VULCAN},
{\it FELIX} 10kJ combining  4 beam lines~\cite{OSAKA},
{\it OMEGA} 30kJ combining 60 beam lines~\cite{OMEGA},
{\it PETAL with quad LMJ} 1.5kJ-80kJ~\cite{PETAL},
{\it LMJ} 1.8MJ combining 240 beam lines~\cite{LMJ}, and
{\it NIF} 1.8MJ combining 192 beam lines~\cite{NIF}.
The quoted numbers above should be regarded as
rough references which can change depending on the operational conditions
and the progress of the on-going projects.
The bottleneck of the currently available laser systems with respect to
the proposed method, {\it the higher the pulse energy, the lower the repetition rate}, 
is going to be improved.
The proposed technique of Coherent Amplification Network (CAN)
\cite{Mourou2012} adopts the coherent addition of highly efficient
fiber lasers which is in principle operational at a higher repetition rate
resolving the heat problem typically seen in the Nd:glass-based lasers.
An international project, International Coherent Amplification Network (ICAN), 
has been launched~\cite{ICAN-HP}.
This is also encouraging for the community of high energy physics aiming at
much higher energy than that of the conventional acceleration technique, 
because the high average power is eventually necessary for high luminosity 
physics even based on the new acceleration scheme; Laser Wakefield Acceleration
(LWFA)~\cite{TT-Dawson}. High-intensity lasers can serve for 
the extension of the traditional course of high energy physics as well as
the novel type of physics as discussed in this paper.
International Center for Zetta- and Exawatt Science and Technology (IZEST)
has been launched toward the integration of high energy physics and 
high-field science~\cite{IZEST-HP}.
These combined efforts are reviewed in our recent article~\cite{TajimaHomma}.

\begin{figure}
\bcent
\includegraphics[width=16.0cm]{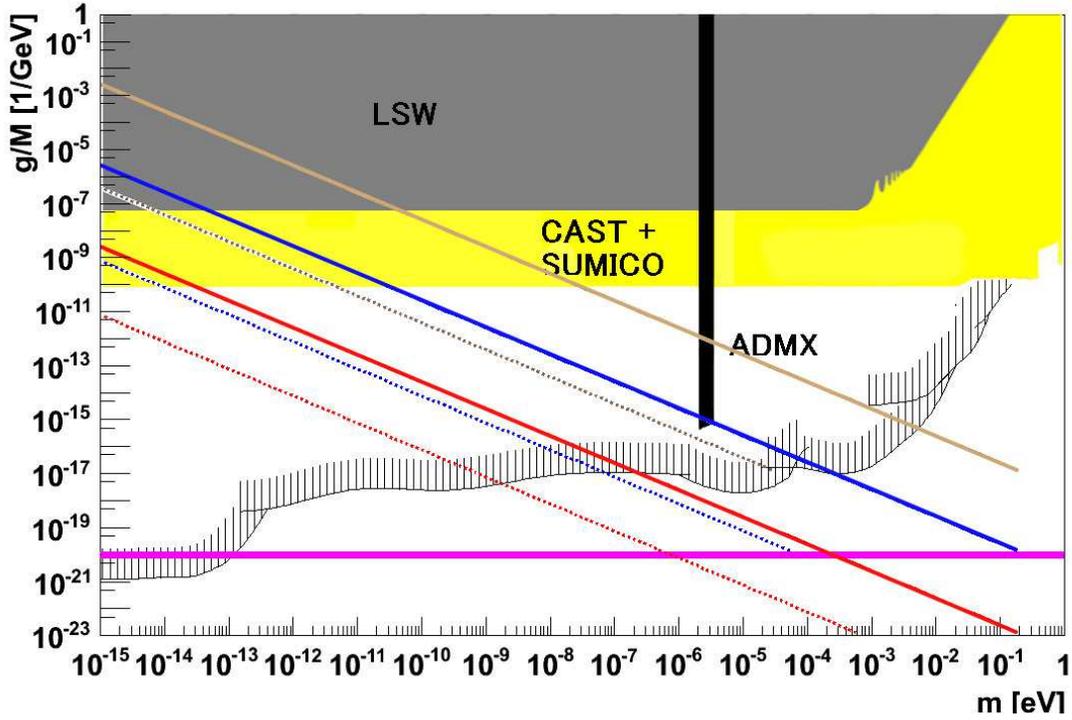}
\caption{
Upper limits of the sensitivities in the $(m, g/M)$ plane by searching
for the four-wave mixing process with a single laser shot.
The brown, blue and red lines indicate the achievable upper limits
with 95\% confidence level by assuming no background process is included,
when no photon with frequency upshit is observed per single shot focusing,
whose laser parameters are summarized in Tab.\ref{Tab1}
based on the formula Eq.(\ref{eq24}) with $C_{mb}=1/2$.
Here ${\cal F}_s = 2\pi$ is used because any ${\cal F}_S$ are on the
same order as those in the axial symmetric case and
the intention of this plot is not in the separation between scalar and
pseudoscalar fields.
The solid and dashed lines are the cases with short and long focal lengths.
The excluded gray, black and yellow domains via the dark field - photon coupling
are quoted from~\cite{AxionReview}. The upper limits 
from the searches for the non-Newtonian force by reinterpreting them based on 
the effective Yukawa interaction between test bodies~\cite{FifthYukawa} 
are also shown with the black shaded curves.
The limit of the gravitational coupling strength is added by assuming
$g\sim{\cal O}(\alpha_{qed})\sim 10^{-2}$ and $M = M_P \sim 10^{18}$~GeV
as an order estimate based on the model~\cite{STTL}.
}
\label{Fig5}
\ecent
\end{figure}

Figure \ref{Fig5} indicates explorable domains in the $(m, g/M)$ plane
by searching for the four-wave mixing process by counting the number
of photons in the frequency-band
$\underline{\omega_3^{'}} \leq \omega_3 \leq \overline{\omega_3^{'}}$,
where ${\cal F}_S = 2\pi$ is used because any ${\cal F}_S$ are on the same
order as those in the axial symmetric case as we discuss in detail
in Appendix and the intention of this plot is not in the separation 
between scalar and pseudoscalar fields.
With the help of the advanced technology,
the frequency-band selection around the optical frequency domain
can be achieved by combining a set of high-quality optical elements
such as prisms, dichroic mirrors and filters
to shut out the non-interacting $1\omega$ and $u\omega$ laser fields
at the downstream of Fig.\ref{Fig1} before the detection of $\omega_3$
with the sensitivity to a single photon.
The single photon detection is not a difficult issue given by any
conventional photomultipliers with the typical gain factor of $\sim 10^6$
with respect to the single photoelectron caused by the incidence of the
single photon especially in the environment where coincidence signals
synchronized with injections of short laser pulses in time are available 
for the rejection of the dark current noise of the photodevice.
%The most difficult aspect of the measurement is, however, 
%in the suppression of the small fraction of the scattering amplitudes 
%from the surface of the prism for the selection of the signal wavelengths 
%by the dispersive feature, which is uncontrollable and some of non-interacting
%photons are accidentally accepted by the photon detector designed to measure
%only the signal wavelengths. However, the commercially available dichroic
%filters can have the discrimination power per single filter expressed as
%the optical density $OD \equiv -\log_{10}(I/I_0) \sim 4 - 5$ with the 
%incident and output photon intensities $I_i$ and $I_o$, respectively.
%Therefore, the multiple dichroic filters can sufficiently suppress the 
%non-interacting photons in the specified spectrum range in advance of 
%the photon detection after the pre-selection by the prism.

The brown, blue, and red lines indicate the achievable upper limits
with 95\% confidence level when no photon in the frequency-band 
$\underline{\omega_3^{'}} \leq \omega_3 \leq \overline{\omega_3^{'}}$
is observed per single shot focusing~\cite{PDGCFL},
whose parameters are summarized in Tab.\ref{Tab1}.
We choose the laser parameters, which heavily depend on the actual system,
as general as possible by considering the anticipated pulse energies
in the existing facilities reviewed above, where the wavelengths of
production and inducing laser beams are assumed to be around 1~$\mu$m,
and the focusing parameters and pulse duration time are chosen so that 
the laser intensity~[W/cm${}^2$] at the surface of the final focusing 
device is lower than the damage threshold typically $10^{13}$~W/cm${}^2$
by more than three orders of magnitude considering the future pulse 
compression option with sub-ps duration. In addition, the
fluence~[J/cm${}^2$] is also required to be lower than the typical 
damage threshold 10J/cm${}^2$ for ns duration. 
The solid and dashed lines are for the short and long 
focal lengths, respectively. This figure provides a baseline to argue 
the single shot sensitivities. Increasing the shot statistics and 
shortening pulse duration time improves the sensitivity. 
These depend on the future development of the high-intensity laser 
technology.

%The parameter is based upon a typical Ti:Sapphire system where
%800nm is produced by pumping laser of 1064nm.
We note that the physical background process from the QED box diagram
is totally negligible as discussed in~\cite{DEptp} essentially due to
smallness of the CMS energies in QPS. 
The generation of high harmonics from residual atoms is 
expected to be a background process by the atomic recombination process
between the ejected electrons and the parent ion after the tunneling
or barrier-suppression ionization by a strong external laser field.
The appearance intensity values are expected
to be $1.5\times 10^{16}$W/cm${}^2$ and $4.0\times 10^{16}$W/cm${}^2$ for
$N^{5+}$ and $O^{6+}$, respectively~\cite{PaulGibbson}.
The vacuum pressure around focal spot, therefore,
should be maintained as low as possible.
The vacuum pump commercially available can achieve
$\sim 10^{-10}$ Pa, where the expected number of atoms per
(100$\mu$m)${}^3$ volume can be below unity.
We can estimate such a background process by requiring
false combinations of linear polarization states of the initial
and final photons in actual measurements. However, for simplicity,
we assumed no background process in order to provide 
the ideal sensitivity curves at this stage.

Filled areas are excluded domains by the other types
of laboratory experiments focusing on the Axion-Like Particle(ALP) 
- photon coupling~\cite{AxionReview} as well as the upper limits
from the searches for non-Newtonian forces by reinterpreting them 
based on the effective Yukawa interaction 
between test bodies~\cite{FifthYukawa}.
The state-of-the-art methods to search for ALP
at terrestrial laboratories by utilizing the two photon-axion coupling are 
represented by LSW(Light-shining-through-walls)\cite{LSW}, 
the solar axion search CAST(CERN Axion Solar Telescope)\cite{CAST} 
and SUMICO(Tokyo Axion Helioscope)\cite{SUMICO}, and 
ADMX (Axion Dark Matter eXpreiment)\cite{ADMX}.
In LSW  a laser pulse together with a static magnetic field produces
ALP and the ALP penetrates an opaque wall thanks to the weakly coupling
nature with matter, and it then regenerates a photon via
coupling to the same static magnetic field located over the wall.
The solar Axion search is similar to LSW,
but different as for the production part.
In the Sun two incoherent photons may produce ALP and the long-lived ALP
penetrates the Sun and the atmosphere in the earth,
and they regenerate photons by coupling to a prepared static magnetic
field on the earth.
ADMX utilizes a microwave cavity immersed in a static magnetic field,
and ALP passing through the cavity can resonantly convert
into real microwave photons.

The Bosonic enhancement is partly utilized in these axion
searches where the axion decay is commonly induced under static 
magnetic field. However, the static magnetic field is 
not in a degenerate state with a narrow momentum range.
Therefore, the enhancement of the decay is limited. The enhancement
of the production rate is also limited because of the broad range
of the CMS energy when choosing two photons for the production of the 
resonance state. We emphasize that the most different aspect of 
our approach is in the field theoretical treatment 
by which we can incorporate the nature of the resonance production and 
decay under the degenerate fields.
This is in contrast to the classical treatment prescribed for the past
axion searches. 
Moreover, the bulk static magnetic field has the limitation
to increase the field strength compared with the recent leap of the
laser energy~\cite{ISMD2011}, where the cutting-edge laser technology
is about to exceed Avogadro's number of photons per laser pulse 
(200kJ $\sim 10^{23}$ optical photons).

\begin{table}[]
\begin{tabular}{c|c|c|c}
\hline\hline
Laser parameters & {\bf Brown lines } & {\bf Blue lines } & {\bf Red lines }\\
\hline\hline
production energy & 200~J & 20~kJ & 2~MJ\\
%production wavelength & 800nm & 800nm & 800nm\\
inducing energy  & 200~J & 20~kJ & 2~MJ\\
%inducing wavelength & 1064nm & 1064nm & 1064nm\\
inducing spectrum width $\delta u$ & 0.1 & 0.1 & 0.1\\
pulse duration time $\tau (= \tau_c = \tau_i)$ & 1~ns & 1~ns & 1~ns\\
diameter at the final focusing mirror $d$ & 40~cm & 80~cm & 800~cm\\
intensity at the final focusing mirror& 
$3.2 \times 10^{8}~$W/cm${}^2$ & $8.0 \times 10^{9}~$W/cm${}^2$ & $8.0 \times 10^{9}~$W/cm${}^2$\\
fluence at the final focusing mirror& $0.32~$J/cm${}^2$ & $8.0~$J/cm${}^2$ & $8.0~$J/cm${}^2$\\
Short focal length $f$ (mass range) & 
1.5~m ($< 0.22$~eV) & 3~m ($< 0.22$~eV) & 30~m ($< 0.22$~eV)\\
$gm/M$ at 95\% C.L. for short focal length&
$2.52 \times 10^{-27}$ & $2.52 \times 10^{-30}$ & $2.52 \times 10^{-33}$\\
Long focal length $f$ (mass range) & 
10km ($< 32$~$\mu$eV) & 10km ($< 65$~$\mu$eV) & 10km ($< 0.65$~meV)\\
$gm/M$ at 95\% C.L. for long focal length&
$3.88 \times 10^{-31}$ & $7.61 \times 10^{-34}$ & $7.56 \times 10^{-36}$\\
\hline \hline
\end{tabular}
\caption{
Laser parameters to estimate the sensitivity to mass-coupling domains
by searching for the four-wave mixing process in the vacuum.
}
\label{Tab1}
\end{table}

\section{Conclusion}
We have shown that the sensitivity to dark fields
by searching for the four-wave mixing process of laser fields 
is expected to be able to reach the sub-eV mass range with 
the coupling strength as weak as that of gravity and even beyond it, 
if the cutting-edge laser technology is properly combined.
Even before reaching extremely high fields, we have many opportunities
to test the light cold Dark Matter candidate by the proposed method.
This high-sensitivity is essentially realized by the Bosonic 
nature of laser fields.

%\section{Fermionic collider vs. coherent Bosonic collider}
As a concluding remark, we emphasize some of features to search for the
four-wave mixing process by comparing them with those in
high-energy colliders as follows. 
As an example, let us remind of the Higgs production 
at LHC as a search for the heavy scalar field.

%First, colliders probe a resonance state by making the CMS energy 
%between accelerated particles closer to the mass of the resonance via 
%the accurately controlled beam momentum. In the case of the hadron collider, 
%in addition to the specification of the beam momentum of the incident hadron, 
%the requirement of the high transverse momentum and the rapidity range of 
%the produced particles restricts the range of the CMS energies 
%between colliding partons effectively.
First, resonance searches in collider experiments are based 
on measurement of the invariant mass distribution of a produced 
resonance state.
In searching for a resonance state in QPS, in contrast, we have the 
unavoidable CMS energy uncertainty originating from the uncertainty 
principle of optical waves compared with the extremely narrow resonance width 
due to the weakly coupling. We are unable to reconstruct the invariant
mass distribution directly, though the interaction probability is still 
affected by the integrated effect over the possible CMS energy uncertainty. 
By observing the appearance or disappearance of the four-wave mixing signal, 
however, one can determine the order of the mass scale from the incident 
wavelength and the collision geometry, a significant difference
from the conventional collider's approach.

Second, in the case of Higgs at LHC,
the dominant production channel is the gluon-gluon fusion process and 
the produced Higgs resonance state decays into two photons, where
both the initial gluons and the final photons are not in degenerate states.  
Therefore, all fields should be treated incoherently, and the decay 
process occurs only via spontaneous processes in the vacuum, {\it i.e.}, 
two photons in the final state are created from 
the pure vacuum state $|0\rangle$.
On the other hand, in the case of photon-photon interactions under 
laser fields, all photons are annihilated into and created from 
the degenerate states. This situation results in
the interaction rate with the cubic dependence on the average number of photons 
included in the laser fields, which is in contrast to the square dependence 
of the number of charged particles in luminosity of 
the Fermionic particle colliders.
The prime mission of the energy frontier of high energy physics is,
of course, to produce new heavy particles, therefore, 
the realization of the high CMS energy is the most important task, while
the sensitivity to weakly coupling fields is sacrificed. 
The dimensionless intensity included in luminosity is proportional 
to the square of the number of charged particles per bunch which is 
typically $\sim 10^{11}$
particles due to the physical limitation by the space-charge effect. 
Even if we could collide them at most 1GHz over 
three years data taking period, the integrated dimensionless intensity
reaches $(10^{11})^2 \cdot 10^9 \cdot 10^8 = 10^{39}$. This indicates that 
it is practically impossible to reach the sensitivity to cross sections 
with coupling including $M^{-1}_P$. 
In contrast, by the proposed method, we can expect the dimensionless
intensity of $(10^{23})^3 = 10^{69}$ even with a single laser shot
including the Avogadro's number of photons.
This manifestly shows how the proposed approach can be sensitive to 
the weakly coupling interactions.

Therefore, in addition to the present most powerful experimental approach
such as heavy boson searches at the high-energy Fermionic collider, 
the proposed coherent Bosonic collider with the inducing mechanism, 
simply speaking, four-wave mixing by focused high-intensity laser fields
opens up a novel opportunity to bridge particle physics and cosmology
in the so far unprobed low-mass and weakly coupling domains
under the controllable laboratory environments.

%\appendix{APPENDIX:Derivation of $\theta_4$ in (\ref{eq20}) from (\ref{eq19})}\\

%From PTP(2.1-3)
%\beqa\label{eq001}
%\omega_3 + \omega_4 = 2\omega
%\eeqa
%\beqa\label{eq002}
%\omega_3\cos\theta_3 + \omega_4\cos\theta_4 = 2\omega\cos\vartheta
%\eeqa
%\beqa\label{eq003}
%\omega_3\sin\theta_3 = \omega_4\sin\theta_4.
%\eeqa

%If we ignore higher order terms more than $\vartheta^2$ and $\theta^2_4$, 
%(\ref{eq19}) is approximated as
%\beqa\label{eq001}
%1-\frac{1}{2}\theta^2_4 \sim 
%\frac{R(2-\frac{1}{2}\vartheta^2)-\vartheta^2}{2R(1-\frac{1}{2}\vartheta^2)}
%= \frac{4R-(R+2)\vartheta^2}{4R-2R\vartheta^2}.
%\eeqa
%From this
%\beqa\label{eq002}
%\theta^2_4 = 2\left(1-\frac{4R-(R+2)\vartheta^2}{4R-2R\vartheta^2}\right)
%= \frac{2-R}{R} \cdot \frac{\vartheta^2}{2-\vartheta^2}
%= \frac{2-R}{R} \cdot \frac{\vartheta^2(2+\vartheta^2)}{4-\vartheta^4}.
%\eeqa
%By taking $\vartheta^4 \sim 0$ 
%\beqa\label{eq003}
%\theta^2_4 \sim \frac{2-R}{2R} \vartheta^2
%\eeqa
%By substituting (\ref{eq17}) into this, we obtain (\ref{eq20}) as follows
%\beqa\label{eq004}
%\theta_4 \sim \sqrt{\frac{2-\frac{u}{2-u}}{2 \frac{u}{2-u}}}\vartheta
%= \sqrt{\frac{4-3u}{2u}}\vartheta.
%\eeqa

\section*{Appendix: Polarization Dependence of Scattering Amplitudes
and Axial Asymmetric Factors ${\cal F}_S$}\
Given the scattering configuration illustrated in Fig.{\ref{Fig3}},
the Lorentz invariant $s$-channel scattering amplitudes
defined in Eq.(\ref{eq_phisigma}) have the following basic form
\beq\label{eqA1}
{\cal M}_S = -(gM^{-1})^2
\frac{{\cal V}^{[1]}_{ab} {\cal V}^{[2]}_{cd}}{(p_1+p_2)^2+m^2},
\eeq
where 
$S\equiv abcd$ with $a,b,c,d = 1$ or $2$, respectively,
denotes a sequence of four-photon polarization states
and $m$ is the mass of scalar or pseudoscalar field.

The vertex factors in the numerator for the case of the
scalar field exchange ($SC$) are defined as
\beqa\label{eqA2}
{\cal V}^{[1]SC}_{ab} \equiv 
<p_2,e_2^{(b)}|\frac{1}{4}F_{\mu\nu}F^{\mu\nu}|p_1,e_1^{(a)}>
= 
\frac{1}{2}<p_2,e_2^{(b)}|F_{\mu\nu}|0><0|F^{\mu\nu}|p_1,e_1^{(a)}>
\nnb\\
{\cal V}^{[2]SC}_{cd} \equiv
<p_4,e_4^{(d)}|\frac{1}{4}F_{\mu\nu}F^{\mu\nu}|p_3,e_3^{(c)}>
=
\frac{1}{2}<p_4,e_4^{(d)}|F_{\mu\nu}|0><0|F^{\mu\nu}|p_3,e_3^{(c)}>,
\eeqa
while these for the case of the pseudoscalar exchange ($PS$) are given by
\beqa\label{eqA3}
{\cal V}^{[1]PS}_{ab} \equiv 
<p_2,e_2^{(b)}|\frac{1}{4}F_{\mu\nu}
\epsilon^{\mu\nu\rho\sigma}F_{\rho\sigma}|p_1,e_1^{(a)}>
= 
\frac{1}{2}<p_2,e_2^{(b)}|F_{\mu\nu}|0><0|
\epsilon^{\mu\nu\rho\sigma}F_{\rho\sigma}|p_1,e_1^{(a)}>
\nnb\\
{\cal V}^{[1]PS}_{cd} \equiv 
<p_4,e_4^{(d)}|\frac{1}{4}F_{\mu\nu}
\epsilon^{\mu\nu\rho\sigma}F_{\rho\sigma}|p_3,e_3^{(c)}>
= 
\frac{1}{2}<p_4,e_4^{(d)}|F_{\mu\nu}|0><0|
\epsilon^{\mu\nu\rho\sigma}F_{\rho\sigma}|p_3,e_3^{(c)}>.
\eeqa

Let us define the polarization vectors and momentum vectors for 
four photons in Fig.\ref{Fig3} as follows:
\beq\label{eqA4}
e_i^{(1)} = (0, 1, 0),
\eeq
\beqa\label{eqA5}
e_1^{(2)} = (-\cos\vartheta, 0, \sin\vartheta), \quad
e_2^{(2)} = (-\cos\vartheta, 0, -\sin\vartheta),
\nnb\\
e_3^{(2)} = (-\cos\theta_3, 0, \sin\theta_3), \quad
e_4^{(2)} = (-\cos\theta_4, 0, -\sin\theta_4),
\nnb\\
p_1=(\omega\sin\vartheta,0,\omega\cos\vartheta; \omega), \quad
p_2=(-\omega\sin\vartheta,0,\omega\cos\vartheta; \omega),
\nnb\\
p_3=(\omega_3\sin\theta_3,0,\omega_3\cos\theta_3; \omega_3), \quad
p_4=(-\omega_4\sin\theta_4,0,\omega_4\cos\theta_4; \omega_4).
\eeqa
Based on these vectors, 
let us summarize basic relations between momenta and polarization vectors
with photon labels $i=1,2,3,4$ as follows
\beq\label{eqA6}
\left(p_i e_j^{(1)} \right) = 0 
\eeq
for the coplanar condition where
the plane determined by $p_1$ and $p_2$ is the same as that of
$p_3$ and $p_4$,
\beqa\label{eqA7}
\left(e_i^{(1)} e_j^{(1)} \right) = 1, \mbox{\quad and \quad }
\left(e_i^{(1)} e_j^{(2)} \right) = 0,
\eeqa
for any pair $i$, $j$, and
\beqa\label{eqA8}
\left(e_i^{(2)} e_j^{(2)} \right) = 1 \mbox{\quad for\quad } i=j, \nnb\\
\left(e_1^{(2)} e_2^{(2)} \right) = \cos2\vartheta,
\left(e_3^{(2)} e_4^{(2)} \right) = \cos(\theta_3+\theta_4)
\equiv\cos\theta_+, \nnb\\
\left(e_1^{(2)} e_3^{(2)} \right) = \cos(\vartheta-\theta_3),
\left(e_2^{(2)} e_4^{(2)} \right) = \cos(\vartheta-\theta_4), \nnb\\
\left(e_1^{(2)} e_4^{(2)} \right) = \cos(\vartheta+\theta_4),
\left(e_2^{(2)} e_3^{(2)} \right) = \cos(\vartheta+\theta_3).
\eeqa
 
We then introduce a clock-wise rotation of the $p_3$-$p_4$ plane from 
the $p_1$-$p_2$ plane defined on the $x-z$ plane by the azimuthal
angle $\varphi$ varying from 0 to $2\pi$ around the $z$-axis in order to
discuss the axial symmetry of the scattering process, 
when polarization vectors are fixed in an experiment. 
The rotated vectors are defined as
\beqa\label{eqA9}
p_3(\varphi) = (\omega_3\sin\theta_3\cos\varphi, 
-\omega_3\sin\theta_3\sin\varphi, \omega_3\cos\theta_3; \omega_3)\nnb\\
p_4(\varphi) = (-\omega_4\sin\theta_4\cos\varphi, 
\omega_4\sin\theta_4\sin\varphi, \omega_4\cos\theta_4; \omega_4),
\eeqa
and
these result in
\beq\label{eqA10}
\left(p_3(\varphi)p_4(\varphi)\right) = 
\omega_3\omega_4(\cos\theta_+-1) = \omega^2(\cos2\vartheta-1)
\eeq
where the last equation is obtained from $(p_1+p_2)^2 = (p_3 + p_4)^2$.
 
With vectors defined above, the vertex factors for the scalar case
are expressed as
\beqa
{\cal V}_{ab}^{[1]SC}&=&(p_1p_2)(e_1^{(a)}e_2^{(b)})-(p_1e_2^{(a)})(p_2e_1^{(b)})\qquad\qquad\quad,\nnb\\
{\cal V}_{cd}^{[2]SC}&=&(p_3(\varphi)p_4(\varphi))(e_3^{(c)}e_4^{(d)})-(p_3(\varphi)e_4^{(c)})(p_4(\varphi)e_3^{(d)}),
\label{eqAscVV}
\eeqa
and these for the pseudoscalar case are expressed as
%%%%%%%%
\beqa
{\cal V}_{ab}^{[1]PS}&=&-\epsilon^{\mu\nu\rho\sigma}p_{1\mu} p_{2\rho} e_{1\nu}^{(a)}e_{2\sigma}^{(b)}\qquad\quad\nnb\\
V_{cd}^{[2]PS}&=&-\epsilon^{\mu\nu\rho\sigma}p_3(\varphi)_\mu p_4(\varphi)_\rho e_{3\nu}^{(c)}e_{4\sigma}^{(d)}.
\label{eqApsVV}
\eeqa

We are now ready to estimate the factor ${\cal F}_S$ included in
the partially integrated cross section in Eq.(\ref{eq4}).
First, we estimate $S=1122$ for the scalar exchange.
From the first of Eq.(\ref{eqAscVV}), we obtain
\beqa\label{eqA11}
{\cal V}^{[1]SC}_{11} = 
\omega^2(\cos2\vartheta-1) \equiv {\cal K}\omega^2.
\eeqa
With
\beqa\label{eqA12}
\left(p_3(\varphi) e_4^{(2)} \right) &=& 
-\omega_3(\sin\theta_3\cos\theta_4\cos\varphi+\cos\theta_3\cos\theta_4)\nnb\\
\left(p_4(\varphi) e_3^{(2)} \right) &=& 
\omega_4(\sin\theta_4\cos\theta_3\cos\varphi+\cos\theta_4\cos\theta_3),
\eeqa
we get
\beqa\label{eqA13}
\left(p_3(\varphi) e_4^{(2)} \right) \left(p_4(\varphi) e_3^{(2)} \right)
= -\omega_3\omega_4\{(1+\cos^2\varphi)\sin\theta_3\sin\theta_4\cos\theta_3\cos\theta_4\nnb\\
+(\sin^2\theta_3\cos^2\theta_4+\sin^2\theta_4\cos^2\theta_3)\cos\varphi
\},
\eeqa
and then the second vertex factor in the second of Eq.(\ref{eqAscVV})
is expressed as
\beqa\label{eqAV22SC}
{\cal V}^{[2]SC}_{22} =
\omega_3\omega_4\{ \cos\theta_+(\cos\theta_+-1)+
(1+\cos^2\varphi)\sin\theta_3\sin\theta_4\cos\theta_3\cos\theta_4\nnb\\
+(\sin^2\theta_3\cos^2\theta_4+\sin^2\theta_4\cos^2\theta_3)\cos\varphi
\}.
\eeqa
For $\varphi=0$, this coincides with $|{\cal V}^{[1]SC}_{11}|$ via
the relation
\beqa\label{eqAV22SC0}
{\cal V}^{[2]SC}_{22} (\varphi=0) &=&\omega_3\omega_4\{\cos\theta_+(\cos\theta_+-1)+\sin^2\theta_+\}\nnb\\
&=& \omega_3\omega_4(1-\cos\theta_+) = \omega^2 (1-\cos2\vartheta)
= -{\cal K}\omega^2.
\eeqa

For a small $\vartheta$ we take the following approximation:
%${\cal K}\sim -2\vartheta^2$,
$\sin\theta_3 \sim \sqrt{{\cal R}}\vartheta$ and
$\sin\theta_4 \sim 1/\sqrt{{\cal R}}\vartheta$ from
Eq.(\ref{eq17}) and (\ref{eq20}),
%and $\omega_3\omega_4 \sim 4{\cal R}\omega^2$ for small $u$.
and this results in 
$\cos\theta_+$ as 
\beqa\label{eqAcosp}
\cos\theta_+ \sim 1-\vartheta^2(1+\hat{{\cal R}}),
\eeqa
with
\beqa\label{eqARhat}
\hat{{\cal R}} \equiv \frac{1}{2}({\cal R}+{\cal R}^{-1}).
\eeqa

We then approximate Eq.(\ref{eqAV22SC}) as
\beqa\label{eqAV22SCAP}
{\cal V}^{[2]SC}_{22} 
%\sim \omega_3\omega_4\left[
%(\cos\theta_+ - 1)\cos\theta_+ 
%+ \vartheta^2\{1+\cos^2\varphi+2\hat{{\cal R}}\cos\varphi\}
%\right]
%\nnb\\
%\sim \omega_3\omega_4\vartheta^2\left(
%-1 - \hat{{\cal R}} + 1+\cos^2\varphi + 2\hat{{\cal R}}\cos\varphi\
%\right)
%\nnb\\
\sim \omega_3\omega_4\vartheta^2\{
\cos^2\varphi + (2\cos\varphi-1)\hat{{\cal R}}
\} 
\equiv \omega_3\omega_4\vartheta^2 F(\varphi).
\eeqa
This is consistent with the approximation of Eq.(\ref{eqAV22SC0})
for a small $\vartheta$
\beqa\label{eqAV22SCAP0}
{\cal V}^{[2]SC}_{22} (\varphi=0) \sim \omega_3\omega_4\vartheta^2
(1 + \hat{{\cal R}})
% \equiv \omega_3\omega_4\vartheta^2 F(0)
.
\eeqa

By taking the square of the factorized second vertex factor,
we then naturally define the factor ${\cal F}_S$ for the scalar case
\beqa\label{eqA15}
{\cal F}^{SC}_{1122} \equiv \int^{2\pi}_{0} F^2(\varphi) d\varphi
\sim
2\pi \left( \frac{3}{8}+3 \hat{{\cal R}}^2 -  \hat{{\cal R}} \right).
\eeqa

Second, let us estimate S=1212 for the pseudoscalar field exchange
as follows. From the first of Eq.(\ref{eqApsVV})
with the vector definitions above, we obtain the first vertex factor as
\beqa
{\cal V}_{12}^{[1]PS}&=&-p_{1\mu}p_{2\rho} \epsilon^{\mu y \rho
\sigma}e_{2\sigma}^{(2)}
=-p_{1\mu}p_{2\rho}\left[ \rule[-.1em]{0em}{1.2em}
\epsilon^{\mu y\rho x}(-\cos\vartheta) +\epsilon^{\mu y\rho
z}(-\sin\vartheta)\right] \nnb\\
&=& p_{2\rho}\left[ \rule[-.1em]{0em}{1.2em}
\left( p_{10}\epsilon^{0y\rho x} +p_{1z}\epsilon^{zy\rho x}
\right)\cos\vartheta
+\left( p_{10}\epsilon^{0y\rho z} +p_{1x}\epsilon^{xy\rho z}
\right)\sin\vartheta
\right] \nnb\\
&=& p_{2\rho}\left[ \rule[-.1em]{0em}{1.2em}
 \left(-\omega \epsilon^{0y\rho x} + \omega\cos\vartheta \epsilon^{zy\rho x}
 \right)\cos\vartheta  +
\left(-\omega \epsilon^{0y\rho z} + \omega\sin\vartheta \epsilon^{xy\rho z}
 \right)\sin\vartheta
\right] \nnb\\
&=& \left[ \rule[-.1em]{0em}{1.2em}
\left( -\omega \epsilon^{0yzx}p_{2z} +\omega\cos\vartheta
\epsilon^{zy0x}p_{20} \right) \cos\vartheta +
\left(  -\omega\epsilon^{0yxz}p_{2x}
+\omega\sin\vartheta\epsilon^{xy0z}p_{2z}  \right)\sin\vartheta\right] \nnb\\
&=& \omega^2\left[ \rule[-.1em]{0em}{1.2em}
\left( -\cos\vartheta +\cos\vartheta \right)\cos\vartheta +\left(
-\sin\vartheta  -\sin\vartheta\right)\sin\vartheta \right]
=-2\omega^2 \sin^2\vartheta.
\label{pss_5}
\eeqa

We also get the second vertex factor
from the second of Eq.(\ref{eqApsVV})
with the vector definitions above as follows
\beqa
{\cal V}_{12}^{[2]PS}&=&-\epsilon^{\mu\nu\rho\sigma}p_{3\mu}p_{4\rho}e_{3\nu}^{(1)}e_{4\sigma}^{(2)}=-\epsilon^{\mu y\rho\sigma}p_{3\mu}p_{4\rho}e_{4\sigma}^{(2)} \nnb\\
&=&-\left( -\epsilon^{\mu y\rho x}\cos\theta_4 - \epsilon^{\mu y\rho z}\sin\theta_4 \right)p_{3\mu}p_{4\rho} \nnb\\
&=&\left[ \rule[-.1em]{0em}{1.1em} \left( \epsilon^{0y\rho x}\cos\theta_4 +\epsilon^{0y\rho z}\sin\theta_4 \right)p_{30} +\left( \epsilon^{zy\rho x} \cos\theta_4 p_{3z}+\epsilon^{xy\rho z}\sin\theta_4p_{3x}  \right)\right]p_{4\rho} \nnb\\
&=& -\omega_3\left( \epsilon^{yzx}p_{4z}\cos\theta_4+\epsilon^{yxz}p_{4x}\sin\theta_4 \right)+4\left( \epsilon^{zy0x}\cos\theta_4p_{3z} +\epsilon^{xy0z}p_{3x}  \right)p_{40} \nnb\\
&=& -\omega_3\omega_4\left[ \rule[-.1em]{0em}{1.1em} \left( \cos^2\theta_4+\sin^2\theta_4\cos\varphi   \right)+\left( -\cos\theta_4\cos\theta_3+\sin\theta_4\sin\theta_3\cos\varphi \right)  \right] \nnb\\
&=& -\omega_3\omega_4\left[ \rule[-.1em]{0em}{1.1em} \cos\theta_4(-\cos\theta_3+\cos\theta_4) +\sin\theta_4(\sin\theta_4+\sin\theta_3)\cos\varphi\right].
\label{azm1_15}
\eeqa

For $\varphi=0$, we find
%%%%%%%%
\beqa
{\cal V}_{12}^{{[2]PS}}(\varphi=0)&=&-\omega_3\omega_4\left( -\cos\theta_3\cos\theta_4 +\cos^2\theta_4 +\sin^2\theta_4+\sin\theta_3\sin\theta_4 \right) \nnb\\
&=&-\omega_3\omega_4\left( 1-\cos\theta_+\right) =-\omega^2\left(1-\cos 2\vartheta \right) = {\cal K}\omega^2.
\label{eqAV12PS0}
\eeqa

If we use the same approximations as the scalar case, 
the second vertex factor is approximated as
\beqa\label{}
{\cal V}_{12}^{{[2]PS}} \sim -\omega_3^2\omega_4^2\vartheta^2 
\{
\check{{\cal R}} + ({\cal R}^{-1}+1) \cos\varphi
\}
\equiv -\omega_3^2\omega_4^2\vartheta^2 G(\varphi),
\eeqa
with
\beq
\check{{\cal R}} \equiv \frac{1}{2}({\cal R}-{\cal R}^{-1}).
\eeq

This is consistent with the approximation of Eq.(\ref{eqAV12PS0})
for a small $\vartheta$
\beqa\label{eqAV12PSAP0}
{\cal V}_{12}^{{[2]PS}}(\varphi=0) 
\sim -\omega_3\omega_4\vartheta^2(1+\hat{{\cal R}}).
\eeqa

Again by taking the square of the second vertex factor,
we then naturally define the factor ${\cal F}_S$ for the pseudoscalar case
\beqa\label{eqA15}
{\cal F}^{PS}_{1212} \equiv \int^{2\pi}_{0} G^2(\varphi) d\varphi
\sim
2\pi \{ \check{{\cal R}}^2+\frac{1}{2}({\cal R}^{-1}+1)^2 \}.
\eeqa

Let us confirm relations for the case of $\varphi = 0$ as follows:
The ratio of the invariant amplitude of the pseudoscalar case to the
scalar case as
\beqa\label{eaAscvsps}
\frac
{{\cal M}^{PS}_{1212}(\varphi=0)}
{{\cal M}^{SC}_{1122}(\varphi=0)}
= \frac
{{\cal V}^{[1]PS}_{12}{\cal V}^{[2]PS}_{12}}
{{\cal V}^{[1]SC}_{11}{\cal V}^{[2]SC}_{22}}
= \frac
{-2\sin^2\vartheta\omega^2 \cdot {\cal K}\omega^2 }
{{\cal K}\omega^2 \cdot -{\cal K}\omega^2}
\sim 1
\eeqa
for a low-mass case with a small $\vartheta$.
The other non-vanishing invariant amplitudes are limited to
$S=1111, 2222, 1122, 2211$ for the scalar
exchange and $S=1212, 1221, 2121, 2112$ for the pseudoscalar case.
These relations can be confirmed by repeating routine
calculations performed above.

%
% ack
%
\section*{Acknowledgments}
K. Homma expresses his deep gratitude to T. Tajima and G. Mourou
for many aspects relevant to this subject. 
He has greatly benefited from  D. Habs, K. Witte, S. Sakabe, and V. Zamfir 
by their valuable suggestions and supports.
He cordially thank Y. Fujii for the long-term discussions 
on the theoretical aspects. 
This work was supported by the Grant-in-Aid for 
Scientific Research no.24654069 from MEXT of Japan.

%
% refs
%


\begin{references}
\bibitem{WMAP}
http://lambda.gsfc.nasa.gov/product/map/current/best\_params.cfm.
\bibitem{DEreview}
L. Amendola and S. Tsujikawa, {\it Dark Energy}, 
Cambirdge University Press (2010).
\bibitem{QE}
R. R. Caldwell, R. Dave, and P. J. Steinhardt, 
Phys. Rev. Lett. {\bf 90}, 1582 (1998); 
L. Wang, R. R. Caldwell, J. P. Ostriker, and P. J. Steinhardt, 
Astrophys. J. {\bf 530}, 17 (2000).
\bibitem{STTL}
Y. Fujii and K. Maeda, {\it The Scalar-Tensor Theory of Gravitation}
Cambridge Univ. Press (2003).
\bibitem{STTtrap}
Y. Fujii, Phys. Lett. B {\bf 660}, 87 (2008), arXiv 0709.2211.
\bibitem{BD}
C. Brans and R. H. Dicke, Phys. Rev. {\bf 124}, 925 (1961).
\bibitem{FujiiShort}
A short summary of \cite{STTL} is available. For example,
see Y. Fujii, arXiv:0908.4324 [astro-ph.CO].
\bibitem{QA}
Y. Nomura, T. Watari and T. Yanagida, Phys. Lett. B {\bf 484}, 103 (2000);
K. Choi, Phys. Rev. D {\bf 62}, 043509 (2000);
J. E. Kim and H. P. Nilles, Phys. Lett. B {\bf 553}, 1 (2003);
L. J. Hall, Y. Nomura and S. J. Oliver, Phys. Rev. Lett. {\bf 95}, 141302 (2005).
\bibitem{Chameleon}
P. Brax, C. van de Bruck, A. C. Davis, J. Khoury and A. Weltman, Phys. Rev. D 70, 123518 (2004).
\bibitem{deVega}
H. J. de Vega, N. G. Sanchez, arXiv:astro-ph/0701212.
\bibitem{YFujii}
Y. Fujii, Nature Phys. Sci. {\bf 234}, 5 (1971).
\bibitem{FifthExp}
E. Fischbach and C. Talmadge. {\it The search for non-Newtonian gravity},
AIP Press, Springer-Verlag, New York (1998).
\bibitem{DEptp}
Y. Fujii and K. Homma,
Prog. Theor. Phys. 126: 531-553 (2011), arXiv:1006.1762 [gr-qc].
\bibitem{PQ} 
R. D. Peccei and H. R. Quinn,
Phys. Rev. Lett. {\bf 38}, 1440 (1977).
\bibitem{AxionReview}
For example, see Figure 2 and section 4 in J.~Jaeckel and A.~Ringwald,
{\it The Low-Energy Frontier of Particle Physics},
Ann. Rev. Nucl. Part. Sci.  {\bf 60}, 405 (2010)
arXiv:1002.0329 [hep-ph].
\bibitem{CDM}
Mark P. Hertzberg, Max Tegmark, and Frank Wilczek,
Phys. Rev. D {\bf 78}, 083507 (2008);
Olivier Wantz and E. P. S. Shellard, Phys. Rev. D {\bf 82}, 123508 (2010).
\bibitem{DEapb}
K. Homma, D. Habs, T. Tajima,
Appl. Phys. B 106:229-240 (2012),
(DOI: 10.1007/s00340-011-4567-3),arXiv:1103.1748 [hep-ph].
\bibitem{Yariv}
Amnon Yariv, {\it Optical Electronics in Modern Communications}
(Oxford University Press, 1997).
\bibitem{Glauber}
R. J. Glauber, Phys. Rev. {\bf 131} (1963), 2766.
\bibitem{FWM}
Sylvie A. J. Druet and Jean-Pierre E. Taran,
Prog. Quant. Electr. Vol.7, pp. 1-72 (1981).
\bibitem{QEDchi3}
F. Moulina and D. Bernardb, Opt. Commun. 164, 137-144 (1999).
\bibitem{QEDfwm}
E. Lundstr\"{o}m et al., Phys. Rev. Lett. 96, 083602 (2006);
J. Lundin et al.,  Phys. Rev. A74, 043821 (2006).
\bibitem{QEDlimit}
D. Bernard et al., Eur. Phys. J. D10, 141 (2000).
\bibitem{ISMD2011}
K. Homma, D. Habs, G. Mourou, H. Ruhl, and T. Tajima,
Prog. Theor. Phys. Suppl. No. 193, (2012);
http://www.extreme-light-infrastructure.eu/;
http://www.int-zest.com/index.html.
\bibitem{PaulGibbson}
P. Gibbson, {\it Short Pulse Laser Interactions with Matter
AN INTRODUCTION} (Imperial College Press, 2005).
\bibitem{FifthYukawa}
Dupays A, Masso E, Redondo J, Rizzo C. Phys. Rev. Lett. 98, 131802 (2007).
\bibitem{LSW}
G. Ruoso et al., Z. Phys. C56, 505 (1992);
R. Cameron et al., Phys. Rev. D47, 3707 (1993);
M. Fouche et al. (BMV Collab.), Phys. Rev. D78, 032013 (2008);
P. Pugnat et al. (OSQAR Collab.), Phys. Rev. D78, 092003 (2008);
A. Chou et al. (GammeV T-969 Collab), Phys. Rev. Lett.  100, 080402 (2008);
A. Afanasev et al. (LIPSS Collab.), Phys. Rev. Lett. 101, 120401 (2008);
K. Ehret et al. (ALPS Collab.), Phys. Lett. B689, 149 (2010).
\bibitem{SUMICO}
S. Moriyama et al., Phys. Lett. B434, 147 (1998);
Y. Inoue et al., Phys. Lett. B536, 18 (2002);
M. Minowa et al., Phys. Lett. B668, 93 (2008). 
\bibitem{CAST}
S. Andriamonje et al. (CAST Collab.), JCAP 0704, 010 (2007);
E. Arik et al. (CAST Collab.), JCAP 0902, 008 (2009);
E. Arik et al. (CAST Collab.), Phys. Rev. Lett. 107, 261302 (2011).
\bibitem{ADMX}
S. Asztalos et al., Phys. Rev. D69, 011101 (2004);
% L. Duffy et al., Phys. Rev. Lett. 95, 091304 (2005);
% S.J. Asztalos et al. (ADMX Collab.), arXiv:0910.5914;
% S.J. Asztalos et al., Nucl. Instrum. Methods A656, 39 (2011);
S.J. Asztalos et al., Phys. Rev. Lett. 104, 041301 (2010);
% Chameleon  G. Rybka et al., Phys. Rev. Lett. 105, 051801 (2010);
% Hidden Sector Photon A. Wagner et al., Phys. Rev. Lett. 105, 171801 (2010).
\bibitem{PDGCFL}
Table 33.3 in
K. Nakamura et al. (Particle Data Group), J. Phys. G 37, 075021 (2010)
and 2011 partial update for the 2012 edition.
\bibitem{VULCAN10PW}
http://www.clf.rl.ac.uk/New+Initiatives/The+Vulcan+10+Petawatt+Project/18344.aspx
\bibitem{APOLLON}
http://www.eli-np.ro/documents/ELI-NP-WhiteBook.pdf
\bibitem{ELI}
http://www.extreme-light-infrastructure.eu/pictures/Grand-Challenges-Meeting-Report-id66.pdf
\bibitem{OSAKA}
http://www.ile.osaka-u.ac.jp/zone1/activities/facilities/spec\_e.html
\bibitem{VULCAN}
http://www.clf.stfc.ac.uk/Facilities/Vulcan/Vulcan+laser/12250.aspx
\bibitem{OMEGA}
http://www.lle.rochester.edu/omega\_facility/omega/
\bibitem{PETAL}
http://int-zest.com/pdf/le-garrec.pdf
\bibitem{LMJ}
http://www-lmj.cea.fr/index-en.htm
\bibitem{NIF}
https://lasers.llnl.gov
\bibitem{Mourou2012}
Mourou, G., Fisch, N., Malkin, V.M. Toroker, Z., Khazanov, E. A., Sergeev,
A. M., Tajima, T., and Le Garrec, B., Opt. Comm. 285, 720 (2012).
\bibitem{ICAN-HP}
https://www.izest.polytechnique.edu/izest-home/ican/ican-94447.kjsp?RF=1332339530225
\bibitem{TT-Dawson}
T. Tajima and J. Dawson, Phy. Rev. Lett. 43, 267 (1979).
\bibitem{TajimaHomma}
T. Tajima and K. Homma, Int. J. Mod. Phys. A vol. 27, No. 25, 1230027 (2012).
\bibitem{IZEST-HP}
http:www.izest.polytechnique.edu
\end{references}
\end{document}